\documentclass[10pt]{article}

\usepackage{amsmath,amssymb,amsfonts,amsthm,amsbsy}
\usepackage{mathtools}
\usepackage{multirow}
\usepackage{color}
\usepackage[a4paper,text={16.8cm,22.4cm}]{geometry}
\usepackage{graphicx}
\usepackage{caption}
\usepackage{slashed}
\usepackage{float}
\usepackage{subcaption}
\usepackage{appendix}
\usepackage{booktabs}
\usepackage{bm}
\usepackage{hyperref}
\usepackage{indentfirst}
\usepackage{orcidlink}
\usepackage[symbol]{footmisc}

\begin{document}

\begin{titlepage}

\begin{center}
{\bf\Large Precision calculations of $B\to K^*$ form factors from SCET sum rules beyond leading-power contributions}
\end{center}
\vspace{0.5cm}

\begin{center}
{\bf 
Jing Gao$\orcidlink{0000-0002-4892-9252}, ^{a,}$ 
\footnote[2]{Email: gao@hiskp.uni-bonn.de}
Ulf-G.~Mei{\ss}ner$\orcidlink{0000-0003-1254-442X},^{a,b,c,}$
\footnote[3]{Email: meissner@hiskp.uni-bonn.de}
Yue-Long Shen$\orcidlink{0000-0002-5740-6154}^{d,}$
\footnote[4]{Email: shenylmeteor@ouc.edu.cn}
and Dong-Hao Li$\orcidlink{0000-0002-6060-7926}^{e,}$
\footnote[1]{Corresponding author, lidh@lzu.edu.cn}
}\\ 
\vspace{0.5cm}
{\sl $^a$ \,Helmholtz-Institut f\"{u}r Strahlen- und Kernphysik and Bethe Center for Theoretical Physics, Universit\"{a}t Bonn, D-53115 Bonn, Germany} \\
{\sl $^b$ \, Institute~for~Advanced~Simulation,~Forschungszentrum~J\"{u}lich,~D-52425~J\"{u}lich,~Germany}\\
{\sl $^c$ \, Peng Huanwu Collaborative Center for Research and Education, International Institute for Interdisciplinary and Frontiers, 
Beihang University, Beijing 100191, China}\\
{\sl $^d$\, College of Physics and Photoelectric Engineering, Ocean University of China, Qingdao 266100, China}\\
{\sl $^e$\, MOE Frontiers Science Center for Rare Isotopes, \\and School of Nuclear Science and Technology, Lanzhou University, Lanzhou 730000, China}

\end{center}
\vspace{0.2cm}

\begin{abstract}
We construct light-cone sum rules (LCSR) for the $B\to K^*$ form factors in the large recoil region using vacuum-to-$B$-meson correlation functions, and systematically calculate subleading-power corrections to these form factors at tree level, including next-to-leading power contributions from the hard-collinear propagator, the subleading effective current $\bar{q}\Gamma[i\slashed{D}_{\perp}/(2m_b)]h_v$, and twist-five/six four-particle higher-twist effects.
By incorporating the available leading-power results at $\mathcal{O}(\alpha_s)$ and the corrections to higher-twist $B$-meson light-cone distribution amplitudes from our previous work, we improve the precision of theoretical predictions for $B\to K^*$ form factors and find that the subleading-power contributions amount to 30\% of the corresponding leading-power results.
Employing the Bourrely-Caprini-Lellouch (BCL) parametrization, we determine the numerical results for $B\to K^*$ form factors across the full kinematic range through a combined fit of LCSR predictions in the large recoil region and lattice QCD results in the small recoil region. Using the newly obtained $B\to K^*$ form factors, we compute the branching fractions for the rare decays $B \to K^* \nu_\ell\bar{\nu}_\ell$ in the Standard Model, obtaining $\mathcal{BR}(\bar{B}^0 \to \bar{K}^{*0} \nu_\ell\bar{\nu}_\ell)=8.09(96)\times 10^{-6}$ and $\mathcal{BR}(\bar{B}^+ \to \bar{K}^{*+} \nu_\ell\bar{\nu}_\ell)=9.95(1.05)\times 10^{-6}$. Additionally, we predict that the longitudinal $K^*$ polarization fraction is $F_L=0.44(4)$. 
\end{abstract}

\vfil

\end{titlepage}

\section{Introduction}
\label{sec:intro}

The semi-leptonic $B$ decays induced by the flavor-changing neutral current (FCNC) and accompanied by clean experimental signals, serve as powerful probes of  physics beyond the Standard Model (SM). 
In the prominent semi-leptonic $b\to s\ell^+\ell^-$ decays, several flavor anomalies have been observed, including a $4.0\sigma$ deviation in the experimentally measured $\mathcal{BR}(B^+\to K^+\mu^+\mu^-)$ compared to the Standard Model predictions in the low $q^2$ region, where $q$ denotes the momentum of the lepton pair, and the discrepancy between the angular observable $P_5'(B \to K^{*0}\mu^+\mu^-)$ measured by the LHCb collaboration and the SM predictions in two $q^2$ bins~\cite{Alguero:2023jeh,LHCb:2020lmf,Capdevila:2023yhq}. 
Notably, the branching ratio of $B\to K \nu_\ell\bar{\nu}_\ell$ reported by the Belle~II collaboration exceeds the Standard Model prediction by 2.7$\sigma$~\cite{Belle-II:2023esi}.
$B$-meson decays with a pair of neutrinos in the final state are one of the cleanest channels in the SM, since the electroweak effects in these processes are under control and the QCD effects are fully encoded in the corresponding hadronic form factors. Meanwhile, the $b\to s\ell^+\ell^-$ decays are affected by various `` non-factorizable ” contributions, including the short-distance hard spectator scattering \cite{Beneke:2001at, Beneke:2004dp}, weak annihilation effects \cite{Lyon:2013gba}, and the power suppressed long-distance quark loop contribution \cite{Beneke:2001at, Khodjamirian:2010vf, Gubernari:2020eft}. 
Studying the $b\to s\nu\bar{\nu}$ process also allows us to distinguish among different $Z'$ models introduced to explain the anomalies in $b\to s \ell^+\ell^-$, or further constrain the Wilson coefficients of high-dimensional operators within the Standard Model effective field theory (SMEFT) \cite{Buras:2014fpa, Allanach:2023uxz, Hou:2024vyw}.
    
In order to make precise theoretical predictions for observables in $B\to K^*\nu_\ell\bar{\nu}_\ell$ decay, precision calculations of $B\to K^*$ form factors are of paramount importance. In the high $q^2$ region, the form factors have been computed using lattice QCD simulations in Refs.~\cite{Horgan:2013hoa,Horgan:2015vla} and extrapolated to the entire kinetic region. In the low $q^2$ region, several QCD based methods have been developed to derive the factorization formulas involved in the heavy-to-light transition processes with the help of the heavy quark expansion. 
At leading power in ${\rm{\Lambda_{QCD}}}/m_Q$, the seven $B\to V$ form factors can be expressed as a product of the four effective operators in soft-collinear effective theory (SCET), the so-called A0-type and B1-type SCET form factors, and corresponding coefficient functions with hard fluctuations~\cite{Beneke:2000wa,Bauer:2002aj,Beneke:2003pa,Beneke:2002ph, Lange:2003pk},
\begin{equation}
\label{QCD-SCET-FF}
\mathcal{F}_i^{B\to V}(n\cdot p)=C_i^{(\rm{A0})}(n\cdot p)\xi_a(n\cdot p)+\int d\tau
C_i^{(\rm{B1})}(\tau,n\cdot p)\Xi_a(\tau,n\cdot p), \qquad (a=\|,\perp),
\end{equation}
where $C_{i}^{(\rm{A0})}$~and~$C_{i}^{(\rm{B1})}$~are the hard functions corresponding to A0-type and B1-type operators, respectively, with their explicit expressions up to~$\mathcal{O}(\alpha_s)$~provided in Appendix~\ref{hard function}~\cite{Hill:2004if,Beneke:2005gs,Bauer:2000yr,Beneke:2004rc}. Specifically, $a=\parallel$ for $\mathcal{F}_i^{B\to V}\in\{\mathcal{A}_0,\mathcal{A}_{12},\mathcal{T}_{23} \}$ and $a=\perp$ for $\mathcal{F}_i^{B\to V}\in
\{\mathcal{V},\mathcal{A}_{1},\mathcal{T}_{1},\mathcal{T}_2 \}$. Due to the endpoint divergences arising in the convolution of the jet functions and the light-cone distribution amplitudes, the soft-collinear factorization of form factors $\xi_a$ cannot be directly accessed. In contrast, the B1-type effective matrix elements can be expressed as the convolution of jet functions and hadronic distribution amplitudes. 

Starting from the vacuum-to-light-meson correlation functions with heavy meson interpolating current, LCSR with light-meson distribution amplitudes has been used to study the 
$B\to V$ form factors up to twist-4 at tree level and to twist-2 at $\mathcal{O}(\alpha_s)$ in Refs.~\cite{Ball:1997rj,Ball:1998kk,Ball:2004rg,Bharucha:2015bzk}. Following the analogous strategies, the light-cone sum rules for $B\to V$ form factors with $B$-meson light-cone distribution amplitudes at tree level were constructed in Ref. \cite{Khodjamirian:2006st} and the subleading-power corrections up to twist-4 at tree level were calculated in Ref.~\cite{Gubernari:2018wyi}. The next-to-leading-logarithmic contribution with SCET sum rules was studied in Ref.~\cite{DeFazio:2007hw}. The power corrections to $B\to V$ form factors from two-particle and three-particle higher-twist $B$-meson LCDAs have been computed in Ref.~\cite{Gao:2019lta}. These computations rely on the universal $B$-meson distribution amplitudes with duality assumption of the light-meson channel and the narrow-width approximation for the vector mesons \cite{Khodjamirian:2023wol, Cheng:2017sfk}. The finite-width effects in the $B\to K^*$ form factors were investigated in Refs.~\cite{Descotes-Genon:2019bud,Descotes-Genon:2023ukb}. 
Compared to QCD factorization, the LCSR approach eliminates the endpoint singularity but introduces a systematic uncertainty due to the quark-hadron duality assumption above a continuum threshold $s_0$, which is used to determine the lowest-lying hadronic parameters.

This work aims to systematically investigate the subleading power effects of $B\to K^*$ form factors in QCD by constructing sum rules with $B$-meson LCDAs, following the approach adopted in Refs.~\cite{Cui:2022zwm,Gao:2021sav,Cui:2023jiw}. 
The subleading power corrections explored in the present work arise from three distinct sources: (I) the power-suppressed terms from the heavy quark expansion of the hard-collinear propagator, (II) the subleading effective current $\bar{q}\Gamma[i\slashed{D}_{\perp}/(2m_b)]h_v$ from the weak current $\bar{q}\Gamma b$, (III) the twist-five and twist-six four-body higher-twist contributions.
By performing combined fits with lattice QCD results in the high $q^2$ region and the improved LCSR form factors in the low $q^2$ region with the BCL parametrization \cite{Okubo:1971my,Bourrely:1980gp,Bourrely:2008za}, we determine the central values and correlation matrix of the BCL $z$-expansion coefficients. 
We then explore the observables in the $\bar{B}\to \bar{K}^{*}\nu_\ell\bar{\nu}_\ell$ process, including the differential branching ratios and the longitudinal $K^*$ polarization fraction.

The organization of the  article is as follows: In Section 2, we present the definitions, notations and leading power effective SCET form factors at $\mathcal{O}(\alpha_s)$. 
In Section 3, we show the computation of various power-suppressed contributions up to twist-six and provides the corresponding $B\to K^*$ form factors in the low $q^2$ region with LCSR. 
In section~4, we apply the BCL parametrization in order to get the $B\to K^*$ form factors in the entire momentum region, and determine the $z$-series expansion coefficients and their correlation matrix by a combined fit of form factors from lattice QCD and LCSR. 
The updated predictions for the branching ratio $\bar{B}\to \bar{K}^{*}\nu_\ell\bar{\nu}_\ell$ and longitudinal $K^*$ polarization fraction are also provided. 
Finally, we discuss our results and future prospects in Section 5. Various technical details are collected in the Appendices.


\section{NLL correction to the \texorpdfstring{$B\to K^*$}{} form factors at leading power }

According to the standard Lorentz decomposition of the bilinear quark currents, the $B\to K^*$ form factors are defined in the standard way \cite{Beneke:2000wa}
\begin{equation}\label{ffQCD}
    \begin{aligned}
    c_V\langle V(p,\varepsilon^*)|\bar{q}\gamma_{\mu}b|\bar{B}(p+q)\rangle&=
 -\frac{2iV(q^2)}{m_B+m_V}\epsilon_{\mu\nu\rho\sigma}\varepsilon^{*\nu}p^\rho q^\sigma,
\\
 c_V\langle V(p,\varepsilon^*)|\bar{q}\gamma_{\mu}\gamma_5 b|\bar{B}(p+q)\rangle&=
 \frac{2m_V\varepsilon^*\cdot q}{q^2}q_{\mu}A_0(q^2)
\\
&+(m_B+m_V)\left(\varepsilon^{*}_{\mu}- \frac{\varepsilon^*\cdot q}{q^2}q_{\mu}\right)A_1(q^2)
\\
&-\frac{\varepsilon^*\cdot q}{m_B+m_V}\left[(2p+q)_\mu-\frac{m^2_B-m^2_V}{q^2}q_\mu\right]A_2(q^2),
\\
 c_V\langle V(p,\varepsilon^*)|\bar{q}i\sigma_{\mu\nu}q^\nu b|\bar{B}(p+q)\rangle&=-
 2iT_1(q^2)\epsilon_{\mu\nu\rho\sigma}\varepsilon^{*\nu}p^\rho q^\sigma,
\\
 c_V\langle V(p,\varepsilon^*)|\bar{q}i\sigma_{\mu\nu}\gamma_5q^\nu b|\bar{B}(p+q)\rangle&=
 T_2(q^2)[(m^2_B-m^2_V)\varepsilon^{*}_{\mu}-(\varepsilon^*\cdot q)(2p+q)_\mu]
\\
 &+T_3(q^2)(\varepsilon^*\cdot q)\left[q_\mu-\frac{q^2}{m^2_B-m^2_V}(2p+q)_\mu\right],
    \end{aligned}
\end{equation}
with the convention $\epsilon_{0123}=-1$. The factor $c_V$ denotes the flavor structure of a vector meson with $c_V=1$ for the $K^*$ meson. Additionally, $m_V$ and $m_B$ denote the mass of $K^*$ meson and $B$ meson, respectively. $p$ and $q$ correspond to the momentum of $K^*$ meson and the momentum transfer of weak current, respectively, with $q=p_B-p=m_Bv-p$.

In the following subsection, the calligraphic form factors $\mathcal{F}_i$ represent the linear combinations of the conventionally defined form factors in Eq.~(\ref{ffQCD}),
\begin{equation}\label{FFQCDSCET}
    \begin{aligned}
        &\mathcal{V}(q^2)=\frac{m_B}{m_B+m_V}V(q^2),&\quad &\mathcal{A}_0(q^2)=\frac{m_V}{E_V}A_0(q^2),&\quad
        &\mathcal{A}_1(q^2)=\frac{m_B+m_V}{2E_V}A_1(q^2),\\
        &\mathcal{A}_{2}(q^2)=\frac{m_B-m_V}{m_B}A_2(q^2),&\quad &\mathcal{T}_1(q^2)=T_1(q^2),&\quad 
        &\mathcal{T}_2(q^2)=\frac{m_B}{2E_V}T_2(q^2),\\
        &\mathcal{A}_{12}(q^2)=\mathcal{A}_1(q^2)-\mathcal{A}_2(q^2),&\quad &\mathcal{T}_{23}(q^2)=\mathcal{T}_2(q^2)-T_3(q^2).&&
    \end{aligned}
\end{equation}
Following the procedure outlined in Refs.~\cite{DeFazio:2007hw, DeFazio:2005dx,  Wang:2015vgv, Lu:2018cfc},
we can construct the vacuum-to-$B$-meson correlation functions as follows:
\begin{equation}
    \begin{aligned}
    {\rm{\Pi}}^{(a)}_{\nu\mu,\parallel}(p,q)=&\int d^4 xe^{ip\cdot x}\langle 0|T\{j_{\nu}(x),\bar{q}(0)\Gamma_\mu ^{(a)}b(0)\}|\bar{B}\rangle,
 \\
 {\rm{\Pi}}^{(a)}_{\nu\delta\mu,\perp}(p,q)=&\int d^4 xe^{ip\cdot x}\langle 0|T\{j_{\nu\delta}(x),\bar{q}(0)\Gamma_{\mu}^{(a)}b(0)\}|\bar{B}\rangle,
    \end{aligned}\label{cft-old}
\end{equation}
where $j_{\nu}(x)=\bar{q}'(x)\gamma_{\nu}q(x)$ and $j_{\nu\delta}(x)=\bar{q}'(x)\gamma_{\nu}\gamma_{\delta\perp}q(x)$ are the interpolating currents corresponding to the longitudinal and transverse polarization vector meson states with momentum $p$, respectively. The superscript $(a)$ denotes  different Dirac structures. We further introduce two light-cone vectors $n_\mu$ and $\bar{n}_\mu$, which satisfy the relations $n\cdot \bar{n}=2$ and $n^2=\bar{n}^2=0$. 
In this work we don't intend to study the  power suppressed contribution arising from  the interpolating currents, therefore we keep the leading-power term of the interpolating currents in our calculations. Subsequently, the correlation functions in Eq.~(\ref{cft-old}) can be expressed as
\begin{equation}
{\rm{\Pi}}^{(a)}_{\nu\mu,\parallel}(p,q)=\bar n_\nu{\rm{\Pi}}^{(a)}_{\mu,\parallel}(p,q),\,\,\,{\rm{\Pi}}^{(a)}_{\nu\delta\mu,\perp}(p,q)=\bar n_\nu{\rm{\Pi}}^{(a)}_{\delta\mu,\perp}(p,q),
\end{equation}
with
\begin{equation}
    \begin{aligned}
    &{\rm{\Pi}}^{(a)}_{\mu,\parallel}(p,q)=\int d^4 xe^{ip\cdot x}\langle 0|T\{\bar{q}'(x)\frac{\slashed{n}}{2}q(x),\bar{q}(0)\Gamma_\mu^{(a)}b(0)\}|\bar{B}(p+q)\rangle,
\\
&{\rm{\Pi}}^{(a)}_{\delta\mu,\perp}(p,q)=\int d^4 xe^{ip\cdot x}\langle 0|T\{\bar{q}'(x)\frac{\slashed{n}}{2}\gamma_{\delta\perp}q(x),\bar{q}(0)\Gamma_\mu^{(a)}b(0)\}|\bar{B}(p+q)\rangle.
    \end{aligned}\label{cft}
\end{equation}
For convenience, we conduct our research  in the rest frame of $B$-meson, which allows us to express the four-velocity vector of $B$ meson as $v_\mu=p_B/m_B=(n_\mu+\bar{n}_\mu)/2$. 
\begin{figure}
\centering
\includegraphics[width=0.35\textwidth]{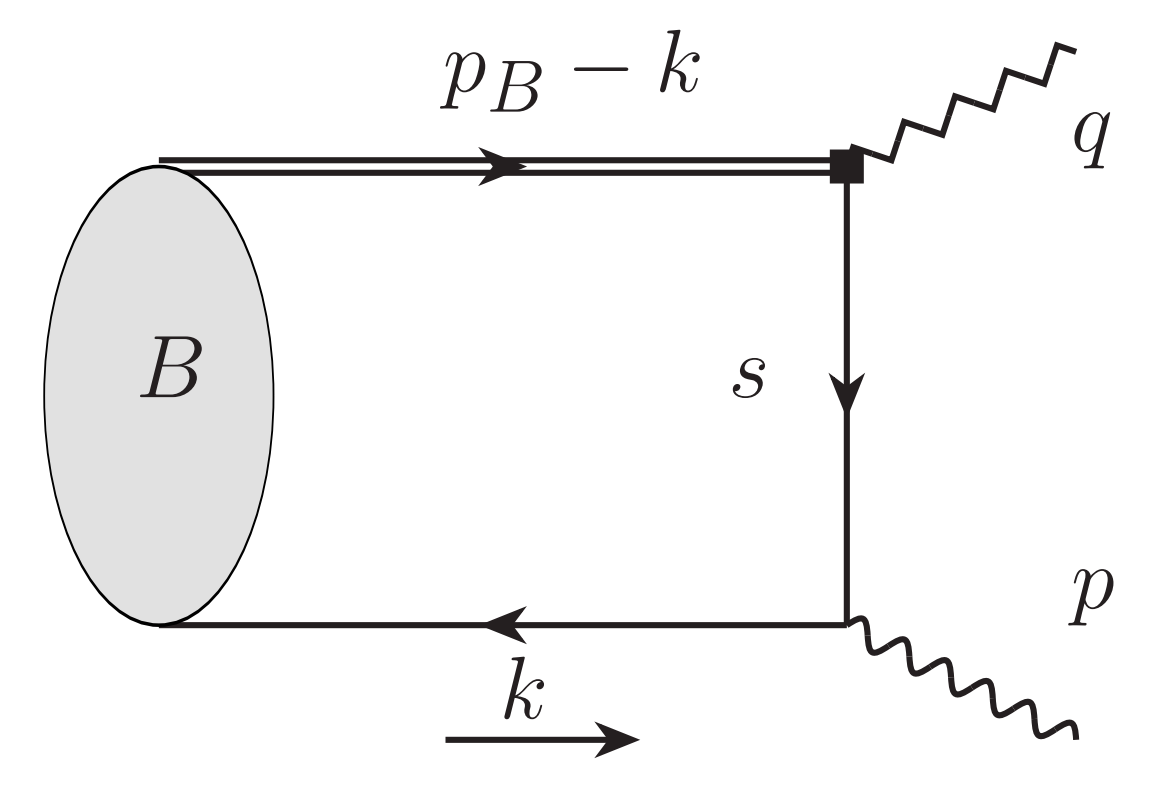}
\caption{Diagrammatic representations of the vacuum-to-$B$-meson correlation functions at tree level, where double line stands for the effective heavy quark field in HQET, the wave line indicates the interpolating current and square box denotes the weak vertex.}
\label{fig:treelevel}
\end{figure}
In addition, the power counting scheme for the momentum of interpolating currents, as well as the masses of  strange quark is assigned to 
\begin{align}
 n\cdot p\sim \mathcal{O}(m_b),\qquad
 |\bar{n}\cdot p|\sim \mathcal{O}({\rm{\Lambda}}_{{\rm{QCD}}}/m_b),\qquad
 m_q\sim m_{q'}\sim \mathcal{O}({\rm{\Lambda}}_{\rm{QCD}}/m_b).
\end{align}
The correlation functions defined in Eq.~(\ref{cft-old}) can be systematically calculated within the framework of SCET.  Since the momentum of the interpolation current is space-like, there is no endpoint singularity in the convolution integral between the perturbation function and the $B$-meson distribution amplitudes, which guarantees factorization of the correlation function.  The two-particle leading-twist $B$-meson LCDA is defined in terms of the nonlocal operators in SCET$_{\rm{II}}$ \cite{Beneke:2000wa,Grozin:1996pq}
\begin{align}
\langle 0|(\bar{q}_sY_s)_\beta(t\bar{n})(Y^{\dagger}_{s}h_v)_\alpha(0)|\bar{B}_v\rangle=-\frac{i\tilde{f}_B(\mu)m_{B}}{4}\left\{\frac{1+\slashed{v}}{2}[2\tilde{\phi}^{+}_{B}(t,\mu)+(\tilde{\phi}^{-}_{B}(t,\mu)-\tilde{\phi}^{+}_{B}(t,\mu))\slashed{\bar{n}}]\gamma_5\right\}_{\alpha\beta},
\end{align}
where
\begin{align}
\tilde{\phi}^{\pm}_{B}(t)\equiv\int_{0}^{\infty}d\omega e^{-i\omega\cdot t}\phi^{\pm}_{B}(\omega).
\end{align} 
The scale-dependent decay constant $\tilde{f}_B(\mu)$ in
HQET is related to the decay constant
$f_B$ in QCD by the following relation \cite{Beneke:2011nf}
\begin{align}
 \tilde{f}_B(\mu)=f_B\bigg[1-\frac{\alpha_s C_F}{4\pi}\bigg(3\ln\frac{m_b}{\mu}-2\bigg)+\mathcal{O}(\alpha^2_s)\bigg].
\end{align}
SCET$_{\rm{II}}$ is an infrared effective theory,  which contains only soft and collinear fields (in this work it is equivalent to HQET since  no collinear field  is taken into account), and it describes all the long distance degrees of freedom. There also exist quark and gluon field at intermediate scale called hard-collinear scale that is described by SCET$_{\rm{I}}$ operators, which deals with the interaction between hard-collinear and soft fields.  The hard-collinear field is at the perturbation region,  and should be integrated out to obtain the jet function.  Therefore, we need to perform the two-step matching process $\rm{QCD}\to \rm{SCET}_{\rm{I}}\to SCET_{\rm{II}}$, then the perturbation functions, including the hard function and jet function, can be obtained step by step. 

Matching the QCD heavy-to-light current to $\rm{SCET}_{\rm{I}}$ operators, which can contribute to the correlation function at leading power, is shown as follows
(see Ref. \cite{Beneke:2005gs} for the explicit expressions of the A-type and B-type SCET operators)
\begin{eqnarray}
(\bar \psi \, \Gamma_i \, Q)(0) &=& \int d {\hat s} \, \sum_{j} \, \tilde{C}_{i j}^{(\rm A0)}(\hat s) \,
O_{j}^{(\rm A0)}(s; 0) + \int d  {\hat s} \, \sum_{j} \, \tilde{C}_{i j \mu}^{(\rm A1)}(\hat s) \,
O_{j}^{(\rm A1) \mu}(s; 0) \nonumber \\
&& +  \int d  {\hat s_1} \, \int d  {\hat s_2} \, \sum_{j} \, \tilde{C}_{i j \mu}^{(\rm B1)}(\hat s_1, \hat s_2) \,
O_{j}^{(\rm B1) \mu}(s_1, s_2; 0) + ...    \,,
\end{eqnarray}
where the hard functions $\tilde{C}_{i j}^{(\rm A0)}(s)$ and $\tilde{C}_{i j \mu}^{(\rm B1)}(\hat s_1, \hat s_2)$ are given in position space,  they could be transformed into momentum space through Fourier transformation, and the corresponding momentum space hard function is written by $C^{(\rm{A0})}_{i}(n\cdot p, \mu)$ and $C^{(\rm{B1})}_{i}(n\cdot p,\tau, \mu)$. Because the hard functions $C^{(\rm{A0})}_{i}$ and $C^{(\rm{B1})}_{i}$ contain the enhanced logarithms $\ln^{n}(m_b/\rm{\Lambda}_{\rm{QCD}})$, 
they should be summed  up to all orders in perturbation theory with NLL and LL accuracy by solving renomalization equation \cite{ Hill:2004if, Beneke:2005gs}. The general solutions to the RG equations are 
\begin{align}
 &C^{(\rm{A0})}_{i}(n\cdot p, \mu)=U_1(n\cdot p, \mu_h, \mu)C^{(\rm{A0})}_{i}(n\cdot p,\mu_h),
 \\
&C^{(\rm{B1})}_{i}(n\cdot p,\tau, \mu)=\exp[-S(n\cdot p, \mu_{h}, \mu)]\int_{0}^{1}d\tau'U_{i}^{(\rm{B1})}(\tau,\tau', \mu_h, \mu)C^{(\rm{B1})}_{i}(n\cdot p,\tau',\mu_h),
\end{align}
where the NLL resummation evolution factor $U_1$ and LL expansion of the $S$ function are detailed in Refs.~\cite{ Beneke:2005gs, Beneke:2011nf}.
The jet function can be obtained by matching the matrix element of the  time-ordered product of $\rm{SCET}_{\rm{I}}$ operators and the SCET$_{\rm I}$ Lagrangian \cite{Beneke:2002ni} to the matrix elements of $\rm{SCET}_{\rm{II}}$ operators. In practical operation, one can just evaluate the Feynman diagrams in Fig.~\ref{fig:treelevel} for tree level result, and the $\mathcal{O}(\alpha_s)$ contribution has been calculated in Ref. \cite{Gao:2019lta}. To extract the form factors of the $B\to K^*$ process, we utilize a dispersion relation and express the partonic correlation function as a dispersion integral \cite{Colangelo:2000dp}
\begin{align}\label{dis 2pt}
 {\rm{\Pi}}(n\cdot p,\bar{n}\cdot p)=\frac{1}{\pi}\int^{\infty}_{0}d\omega' \frac{{\rm{Im}} \,{\rm{\Pi}}(n\cdot p,\omega')}{\omega'-\bar{n}\cdot p-i\epsilon}.
\end{align}
At the hadronic level, the correlation function with different interpolating currents can be expressed by the following formulas
\allowdisplaybreaks
\begin{align}\label{hcf}
    {\rm{\Pi}}_{\mu, \parallel}^{(\rm{V-A})}(p, q)= &  \frac{f^{\|}_{V} m_{V}}{m_{V}^{2} / n \cdot p-\bar{n} \cdot p-i 0}\left(\frac{n \cdot p}{2 m_{V}}\right)^{2}\bigg\{\frac { m _ { B } } { m _ { B } - n \cdot p } n _ { \mu } \bigg[\left(-\frac{2 m_{V}}{n \cdot p} A_{0}\left(q^{2}\right)\right)
    \nonumber
\\
+&\bigg(\frac{m_{B}+m_{V}}{n \cdot p} A_{1}\left(q^{2}\right)
 -\frac{m_{B}-m_{V}}{m_{B}} A_{2}\left(q^{2}\right)\bigg)\bigg]
 \nonumber
\\
 -&\bar{n}_{\mu}\bigg[\bigg(\frac{2 m_{V}}{n \cdot p} A_{0}(q^{2})\bigg)
 +\left(\frac{m_{B}+m_{V}}{n \cdot p} A_{1}\left(q^{2}\right)-\frac{m_{B}-m_{V}}{m_{B}} A_{2}\left(q^{2}\right)\right)\bigg]\bigg\} 
 \nonumber
\\
 +&\int d \omega^{\prime} \frac{1}{\omega^{\prime}-\bar{n} \cdot p-i 0}\left[n_{\mu} \varrho_{n, \|}^{(\rm{V-A})}\left(\omega^{\prime}, n \cdot p\right)+\bar{n}_{\mu} \varrho_{\bar{n}, \|}^{(\rm{V-A})}\left(\omega^{\prime}, n \cdot p\right)\right], 
 \\
{\rm{\Pi}}_{\delta \mu, \perp}^{(\rm{V-A})}(p, q)= & -\frac{1}{2} \frac{f^{\perp}_{V} n \cdot p}{m_{V}^{2} / n \cdot p-\bar{n} \cdot p-i 0}
\nonumber
\\
\times&\left[g^{\perp}_{\delta \mu }\left(\frac{m_{B}+m_{V}}{n \cdot p} A_{1}\left(q^{2}\right)\right)+i \epsilon^{\perp}_{\delta \mu }\left(\frac{m_{B}}{m_{B}+m_{V}} V\left(q^{2}\right)\right)\right] 
\nonumber
\\
+&\int d \omega^{\prime} \frac{1}{\omega^{\prime}-\bar{n} \cdot p-i 0}\left[g^{\perp}_{\delta \mu } \varrho_{\perp, A_{1}}^{(\rm{V-A})}\left(\omega^{\prime}, n \cdot p\right)+i \epsilon^{\perp}_{\delta \mu} \varrho_{\perp, V}^{(\rm{V-A})}\left(\omega^{\prime}, n \cdot p\right)\right],
\\
{\rm{\Pi}}_{\mu, \|}^{(\rm{T+\tilde{T}})}(p, q)= & \frac{1}{2} \frac{f^{ \|}_{V} m_{V}}{m_{V}^{2} / n \cdot p-\bar{n} \cdot p-i 0}\left(\frac{n \cdot p}{2 m_{V}}\right)^{2}
\nonumber
\\
\times&\left[n \cdot q \bar{n}_{\mu}-\bar{n} \cdot q n_{\mu}\right]\left[\frac{m_{B}}{n \cdot p} T_{2}\left(q^{2}\right)-T_{3}\left(q^{2}\right)\right] 
\nonumber
\\
+&\int d \omega^{\prime} \frac{1}{\omega^{\prime}-\bar{n} \cdot p-i 0}\left[n \cdot q \bar{n}_{\mu}-\bar{n} \cdot q n_{\mu}\right] \varrho_{\|}^{(\rm{T+\tilde{T}})}\left(\omega^{\prime}, n \cdot p\right), 
\\
{\rm{\Pi}}_{\delta \mu, \perp}^{(\rm{T+\tilde{T}})}(p, q)= & \frac{1}{2} \frac{f^{\perp}_{V} n \cdot p m_{B}}{m_{V}^{2} / n \cdot p-\bar{n} \cdot p-i 0}\left[g^{\perp}_{\delta \mu }\left(\frac{m_{B}}{n \cdot p} T_{2}\left(q^{2}\right)\right)+i \epsilon^{\perp}_{\delta \mu } T_{1}\left(q^{2}\right)\right] 
\nonumber
\\
+&\int d \omega^{\prime} \frac{1}{\omega^{\prime}-\bar{n} \cdot p-i 0}\left[g^{\perp}_{\delta \mu } \varrho_{\perp, T_{2}}^{(\rm{T+\tilde{T}})}\left(\omega^{\prime}, n \cdot p\right)+i \epsilon^{\perp}_{\delta \mu } \varrho_{\perp, T_{1}}^{(\rm{T+\tilde{T}})}\left(\omega^{\prime}, n \cdot p\right)\right],\label{hcf1}
\end{align}
where the definitions of decay constants for the longitudinal and transverse $K^*$ meson are given by Ref. \cite{DeFazio:2007hw}
\begin{equation}
    \begin{aligned}
    \langle 0|\bar{q'}(x)\frac{\slashed{n}}{2}q(x)|V(p,\varepsilon)\rangle=&
\frac{i}{2}f^{\parallel}_{V}m_V n\cdot\varepsilon,
\\
\langle 0|\bar{q'}(x)\frac{\slashed{n}}{2}\gamma^{\perp}_{\delta}q(x)|V(p,\varepsilon)\rangle=&
-n\cdot p\frac{i}{2}f^{\perp}_{V}\varepsilon^{\perp}_\delta(p).
    \end{aligned}
\end{equation}
At leading power, the large recoil symmetry reduce the seven $B \to V$ form factors to two, namely, $\xi_{\parallel}(n\cdot p)$ and  $\xi_{\perp}(n\cdot p)$.  The relation between the QCD form factors in the above equation and the SCET form factors  $\xi_{\parallel, \perp}(\bar n \cdot p)$ can be found in Ref. \cite{Gao:2019lta}. Taking advantage of these relations, all the correlation functions can be expressed in terms of the SCET form factors. We then apply quark-hadron duality and Borel transformation to both the partonic and hadronic correlation functions in the SCET representation to eliminate the continuum and resonance contributions, thereby reducing the uncertainties from the duality ansatz. We finally obtain the four effective SCET form factors at $\mathcal{O}(\alpha_s)$ \cite{Gao:2019lta}
\allowdisplaybreaks
\begin{align}
  \xi_{\parallel,\rm{NLL}}(n\cdot p)=& \frac{U_2(\mu_{h2},\mu)\tilde{f}_B(\mu_{h2})}{f^{\parallel}_V}
 \frac{2m_Bm_V}{(n\cdot p)^2}
 \nonumber\\
 \times&\int^{\omega^\parallel_s}_{0}d\omega'
 \exp\bigg[-\frac{n\cdot p\omega'-m^{2}_V}{n\cdot p\omega_M}\bigg][\phi^{-}_{B,\text{eff}}(\omega',\mu)+\phi^{+}_{B,m}(\omega',\mu)],
 \\
  \xi_{\perp,\rm{NLL}}(n\cdot p)=&\frac{U_2(\mu_{h2},\mu)\tilde{f}_B(\mu_{h2})}{f^{\perp}_V(\nu)}
 \frac{m_B}{n\cdot p}
 \nonumber\\
 \times&\int^{\omega^\perp_s}_{0}d\omega'
 \exp\bigg[-\frac{n\cdot p\omega'-m^{2}_V}{n\cdot p\omega_M}\bigg]\tilde{\phi}^{-}_{B,\text{eff}}(\omega',\mu,\nu),
 \\
 \Xi_{\parallel,\rm{NLL}}(n\cdot p)=&-\frac{\alpha_sC_F}{\pi}\frac{U_2(\mu_{h2},\mu)\tilde{f}_B(\mu_{h2})}{f^{\parallel}_V}
 \frac{m_Bm_V}{n\cdot p m_b}[(1-\tau)\theta(\tau)\theta(1-\tau)]
 \nonumber\\
 \times&\int^{\omega^\parallel_s}_{0}d\omega'
 \exp\bigg[-\frac{n\cdot p\omega'-m^{2}_V}{n\cdot p\omega_M}\bigg]
 \int^{\infty}_{\omega'}d\omega\frac{\phi^{+}_{B}(\omega,\mu)}{\omega},
 \\
  \Xi_{\perp,\rm{NLL}}(n\cdot p)=&-\frac{\alpha_sC_F}{2\pi}\frac{U_2(\mu_{h2},\mu)\tilde{f}_B(\mu_{h2})}{f^{\perp}_V(\nu)}
 \frac{m_B}{m_b}[(1-\tau)\theta(\tau)\theta(1-\tau)]
 \nonumber\\
 \times&\int^{\omega^\perp_s}_{0}d\omega'
 \exp\bigg[-\frac{n\cdot p\omega'-m^{2}_V}{n\cdot p\omega_M}\bigg]
 \int^{\infty}_{\omega'}d\omega\frac{\phi^{+}_{B}(\omega,\mu)}{\omega},
\end{align}
where $\omega^{\parallel, \perp}_s=s^{\parallel, \perp}_0/n\cdot p$ and $\omega_M=M^2/n\cdot p$ denote
the effective threshold and Borel mass, respectively, which are two fundamental inputs of LCSR.
$\mu$ and $\nu$ correspond to the factorization scale and renormalization scale, respectively. Additionally, the effective $B$-meson distribution amplitudes are introduced
to describe both the hard-collinear and soft fluctuations
\cite{Gao:2019lta, Wang:2015vgv, Lu:2018cfc}, as given in Appendix~\ref{effective-BDA}.
The quark mass contributions in SCET have been investigated in Ref. \cite{Leibovich:2003jd} and the leading-power spectator-quark mass corrections to $B_{(s)}$-meson decay form factors at one-loop accuracy have been calculated in Refs. \cite{Cui:2022zwm, Cui:2023jiw}. In this work, we do not include the spectator-quark mass corrections to the effective SCET form factors at one-loop accuracy. The omission is justified because the spectator quark in $B\to K^*$ decay process is either a $u$-quark or $d$-quark, whose mass-induced corrections are significantly suppressed compared to those from the s-quark in $B_{s}$-meson decays. We will estimate the spectator-quark correction to the $B\to K^*$ form factors with the SCET sum rules in our future work.

\section{Subleading-power contributions}\label{sec 3}
In this section, we investigate the power corrections arising from various sources to the $B\to K^*$ form factors within the LCSR approach. Utilizing the equations of motion in HQET and the factorization formula of correlation functions at subleading power, we ultimately obtain the tree-level power corrections to the $B\to K^*$ form factors in the large hadronic recoil region and analyze the scaling behavior of these form factors.

\subsection{Higher-twist \texorpdfstring{$B$}{}-meson LCDAs contribution}
\begin{figure}[htb]
\centering
\includegraphics[width=0.70\textwidth]{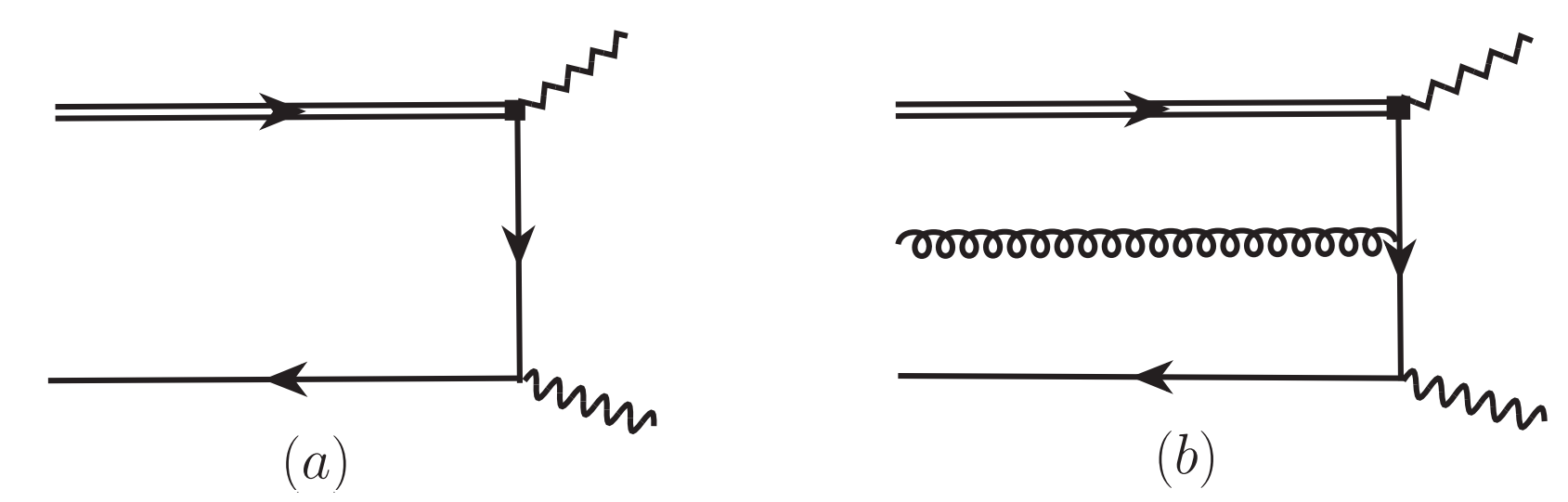}
\caption{Diagrammatic representations of two-particle (left) and three-particle (right) corrections to the vacuum-to-$B$-meson correlation functions at tree level, where the square box indicates the weak vertex and wavy line represents the interpolating current.}
\label{fig:HT}
\end{figure}
The contributions from the higher-twist B-meson LCDAs of two-particle and three-particle are shown in Fig. \ref{fig:HT}. 
In order to define  higher-twist B-meson LCDAs, the general parameterization of the vacuum-to-$B$-meson matrix element of three-body HQET operator is given by Ref.~\cite{Braun:2017liq}
\begin{flalign}\label{3pLCDA}
 &\langle 0|\bar{q}_\alpha(z_1\bar{n})g_sG^{\mu\nu}(z_2\bar{n})h_{v\beta}|0\rangle=
 \frac{\tilde{f}_B(\mu)m_{B}}{4}[(1+\slashed{v})\{(v_\mu\gamma_\nu-v_\nu\gamma_\mu)\bigg[\Psi_A(z_1,z_2,\mu)-\Psi_V(z_1,z_2,\mu)\bigg]
 \nonumber\\
 &-i\sigma_{\mu\nu}\Psi_V(z_1,z_2,\mu)-(\bar{n}_\mu v_\nu-\bar{n}_\nu v_\mu)X_A(z_1,z_2,\mu)
 +(\bar{n}_\mu \gamma_\nu-\bar{n}_\nu \gamma_\mu)\bigg[W(z_1,z_2,\mu)+Y_A(z_1,z_2,\mu)\bigg]
\nonumber\\
&+i\epsilon_{\mu\nu\alpha\beta}\bar{n}^\alpha v^\beta\gamma_5\tilde{X}_A(z_1,z_2,\mu)-i\epsilon_{\mu\nu\alpha\beta}\bar{n}^\alpha \gamma^\beta\gamma_5\tilde{Y}_A(z_1,z_2,\mu)
\nonumber\\
&-(\bar{n}_\mu v_\nu-\bar{n}_\nu v_\mu)\slashed{\bar{n}}W(z_1,z_2,\mu)+
(\bar{n}_\mu \gamma_\nu-\bar{n}_\nu \gamma_\mu)\slashed{\bar{n}}Z(z_1,z_2,\mu)\}\gamma_5]_{\beta\alpha},
\end{flalign}
where~$\epsilon_{0123}=-1$, and we also introduce three-particle HQET distribution amplitudes of definite collinear twist as follows
\begin{equation}
    \begin{aligned}
    &\Phi_3=\Psi_A-\Psi_V,&
    \Phi_4&=\Psi_A+\Psi_V,
\\
 &\Psi_4=\Psi_A+X_A,&
 \tilde{\Psi}_4&=\Psi_V-\tilde{X}_A,
\\
 &\tilde{\Phi}_5=\Psi_A+\Psi_V+2Y_A-2\tilde{Y}_A+2W,&
 \Psi_5&=-\Psi_A+X_A-2Y_A,
\\
 &\tilde{\Psi}_5=-\Psi_V-\tilde{X}_A+2\tilde{Y}_A,&
 \Phi_6&=\Psi_A-\Psi_V+2Y_A+2W+2\tilde{Y}_A-4Z.
    \end{aligned}
\end{equation}
To facilitate the calculation of the contributions from high-twist LCDAs more effectively, we introduce the light-cone expansion of the quark propagator within the background gluon field. The relevant definition for this expansion can be located in the Ref.~\cite{Balitsky:1987bk}.
\begin{align}\label{3pp} 
 \langle 0|T\{\bar{q}(x),q(0)\}|0\rangle\supset ig_s\int \frac{d^4 l}{(2\pi)^4} e^{-il\cdot x}\int^{1}_{0}du
 \bigg[\frac{ux_\mu\gamma_\nu}{l^2-m_{q}^2}-\frac{(\slashed{l}+m_q)\sigma_{\mu\nu}}{2(l^2-m_{q}^2)^2}\bigg]G^{\mu\nu}(ux),
\end{align}
First, we evaluate the correlation function ${\rm{\Pi}}^{(a)}_\parallel(p,q)$ in Eq.~(\ref{cft}), which corresponds to the longitudinally polarized vector meson. Contracting the quark fields $q(x)$ and $\bar q(0)$ (leads to the propagator in Eq.~(\ref{3pp})),  taking advantage of  the three-particle higher-twist $B$-meson LCDA in Eq.~(\ref{3pLCDA}), the three-particle higher-twist correction to the correlation function can be
written as
\begin{equation}
    {\rm{\Pi}}^{(a),\text{3PHT}}_{\parallel,\text{NLP}}(p,q)=\frac{\tilde{f}_B(\mu)m_{B}}{2n\cdot p}\left[
    \Gamma_\parallel^{(a)}
    {\rm{\Pi}}^{\text{3PHT}}_{\parallel,\text{NLP}}+
    \tilde{\Gamma}_\parallel^{(a)}\tilde{{\rm{\Pi}}}^{\text{3PHT}}_{\parallel,\text{NLP}}\right],
\end{equation}
with
\begin{equation}
\begin{aligned}
{\rm{\Pi}}^{\text{3PHT}}_{\parallel,\text{NLP}}&=\int^{\infty}_{0}d\omega_1\int^{\infty}_{0}d\omega_2\int^{1}_{0}du\frac{-2\bar{u}\Phi_4(\omega_1,\omega_2)}{(\bar{n}\cdot p-\omega_1-u\omega_2-\omega_q)^2}+\frac{m_q}{n\cdot p}\frac{\Psi_5(\omega_1,\omega_2)-\tilde{\Psi}_5(\omega_1,\omega_2)}{(\bar{n}\cdot p-\omega_1-u\omega_2-\omega_q)^2},
    \\
    \tilde{{\rm{\Pi}}}^{\text{3PHT}}_{\parallel,\text{NLP}}&=
    \int^{\infty}_{0}d\omega_1\int^{\infty}_{0}d\omega_2\int^{1}_{0}du
    \frac{\tilde{\Psi}_5(\omega_1,\omega_2)-(2u-1)\Phi_5(\omega_1,\omega_2)}{(\bar{n}\cdot p-\omega_1-u\omega_2-\omega_q)^2}-
    \frac{m_q}{n\cdot p}
    \frac{2\tilde{\Phi}_6(\omega_1,\omega_2)}
    {(\bar{n}\cdot p-\omega_1-u\omega_2-\omega_q)^2},
\end{aligned}
\end{equation}
where the factors $\Gamma^{(a)}_\parallel$ and $\tilde{\Gamma}^{(a)}_\parallel$ take the form
\begin{equation}
    \left\{\Gamma^{({\rm{V-A}})}_\parallel,\Gamma^{({\rm{T+\tilde{T}}})}_\parallel\right\}\in\left\{n_\mu,\bar{n}_\mu\frac{n\cdot q}{2}-n_\mu\frac{\bar{n}\cdot q}{2}\right\},\,\,
    \left\{\tilde{\Gamma}^{({\rm{V-A}})}_\parallel,\Gamma^{({\rm{T+\tilde{T}}})}_\parallel\right\}\in\left\{\bar{n}_\mu,n_\mu\frac{\bar{n}\cdot q}{2}-\bar{n}_\mu\frac{n\cdot q}{2}\right\},
\end{equation}
and $a\in\{\text{V}-\text{A},{\rm{T+\tilde{T}}}\}$ denotes the different Dirac structures $\gamma_{\mu}(1-\gamma_5)$~and~$i\sigma_{\mu\nu}(1+\gamma_5)q^{\nu}$ of the heavy-to-light weak current, respectively. 
Additionally, we set $\omega_q=m_{q}^2/n\cdot p$ for brevity.

Along the same lines, the three-particle higher-twist correction to the correlation function ${\rm{\Pi}}^{(a)}_\perp(p,q)$ in Eq.~(\ref{cft}), which corresponds to a transversely polarized vector meson, can be expressed as 
\begin{equation}
\begin{aligned}
    {\rm{\Pi}}^{(a),\text{3PHT}}_{\perp,\text{NLP}}(p,q)=\frac{\tilde{f}_B(\mu)m_{B}}{2n\cdot p}
    \int^{\infty}_{0}d\omega_1\int^{\infty}_{0}d\omega_2\int^{1}_{0}du &\left[\Gamma_\perp^{(a)}\frac{(1-2u)\Psi_5(\omega_1,\omega_2)-\tilde{\Psi}_5(\omega_1,\omega_2)}{(\bar{n}\cdot p-\omega_1-u\omega_2-\omega_q)^2}\right.\\
   & \,-\left.
    \tilde{\Gamma}_\perp^{(a)}\frac{m_q}{n\cdot p}
    \frac{\Psi_5(\omega_1,\omega_2)+\tilde{\Psi}_5(\omega_1,\omega_2)}
    {(\bar{n}\cdot p-\omega_1-u\omega_2-\omega_q)^2}
    \right]
\end{aligned}
\end{equation}
by taking $\epsilon^{\perp}_{\delta\mu}\equiv\frac{1}{2}\epsilon_{\delta\mu\rho\sigma}n^{\rho}\bar{n}^{\sigma}$ and
\begin{equation}
    \begin{aligned}
    \left\{\Gamma^{({\rm{V-A}})}_\perp,\Gamma^{({\rm{T+\tilde{T}}})}_\perp\right\}&\in\{g_{\delta\mu}^\perp+i\epsilon_{\delta\mu},g_{\delta\mu}^\perp-i\epsilon_{\delta\mu}^\perp\},\\
    \left\{\tilde{\Gamma}^{({\rm{V-A}})}_\perp,\tilde{\Gamma}^{({\rm{T+\tilde{T}}})}_\perp\right\}&\in-\{\bar{n}\cdot q(g_{\delta\mu}^\perp+i\epsilon_{\delta\mu}),n\cdot q(g_{\delta\mu}^\perp-i\epsilon_{\delta\mu}^\perp)\}.
    \end{aligned}
\end{equation}
In addition, the two-particle higher-twist $B$-meson LCDAs, for example~$g^{\pm}_{B}(\omega)$, also generate subleading-power contributions. The off-light-cone corrections to the renormalized two-body non-local HQET matrix element at  $\mathcal{O}(x^2)$~accuracy is given by Ref.~\cite{Braun:2017liq}
\begin{align}\label{2pLCDA}
 \langle 0|(\bar{q}_sY_s)_\beta(x)(Y^\dagger_sh_{v})_\alpha|\bar{B}_v\rangle&=
 -i\frac{\tilde{f}_B(\mu)m_{B}}{4}\int^{\infty}_{0}d\omega e^{-i\omega v\cdot x}\left[\frac{1+\slashed{v}}{2}\left\{2\bigg[\phi^+_B(\omega,\mu)+x^2g^+_B(\omega,\mu)\right]
 \right.\nonumber \\
 &\left.-\frac{\slashed{x}}{v\cdot x}\left[(\phi^+_B(\omega,\mu)-\phi^-_B(\omega,\mu))+x^2(g^+_B(\omega,\mu)-g^-_B(\omega,\mu))\bigg]\right\}\gamma_5\right]_{\alpha\beta},
\end{align}  
where the higher-twist LCDA $g^-_B(\omega,\mu)$ can be decomposed into the ``genuine'' twist-five three-particle LCDA 
${\rm{\Psi}}_5(\omega_1,\omega_2,\mu)$ and the lower-twist ``Wandzura-Wilczek'' two-particle LCDA $\hat{g}^{-}_{B}(\omega,\mu)$
\begin{align}
 g^{-}_{B}(x)=\hat{g}^{-}_{B}(x)-\frac{1}{2}\int^{1}_{0}du \bar{u}\Psi_{5}(x,ux).
\end{align}
After inserting the two-particle higher-twist $B$-meson LCDAs given in Eq.~(\ref{2pLCDA}) into the correlation functions, we obtain the factorization formulas for the two-particle higher-twist contributions
\begin{equation}
\begin{aligned}
    {\rm{\Pi}}^{(a),\text{2PHT}}_{\parallel, \perp}(p,q)=&\frac{\tilde{f}_B(\mu)m_{B}}{2n\cdot p}\Gamma_{\parallel, \perp}^{(a)}\left\{\int^{\infty}_{0}d\omega\frac{4\hat{g}^-_B(\omega,\mu)}{(\bar{n}\cdot p-\omega)^2} \right.\\
    &\left.
-\int^{\infty}_{0}d\omega_1\int^{\infty}_{0}d\omega_2\int^{1}_{0}du
 \frac{2\bar{u}\Psi_5(\omega_1,\omega_2)}{(\bar{n}\cdot p-\omega_1-u\omega_2)^2}\right\},
\end{aligned}
\end{equation}
with $\Gamma_\parallel^{(a)}\in\{\bar{n}_\mu,n_\mu\frac{\bar{n}\cdot q}{2}-\bar{n}_\mu \frac{n\cdot q}{2} \}$ and $\Gamma_\perp^{(a)}\in\{g_{\delta\mu}^\perp+i\epsilon_{\delta\mu}^\perp,-\bar{n}\cdot q (g_{\delta\mu}^\perp+i\epsilon_{\delta\mu}^\perp)\}$ for $a\in\{\text{V}-\text{A},{\text{T}+\tilde{\text{T}}}\}$.

Summing up the two-particle and three-particle higher-twist contributions,  
we obtain the higher-twist corrections to the correlation functions ${\rm{\Pi}}_{\parallel,\perp}^{(a)}(p,q)$ at tree level. We then implement the dispersion relation to the correlation functions at the partonic level and apply the quark-hadron duality ansatz and Borel transformation. This procedure yields the following sum rules for the higher-twist corrections to the $B\to K^*$ form factors \cite{Gao:2019lta}
\begin{equation}
    \begin{aligned}
        f_V^\perp & \exp\left[-\frac{m_{V}^{2}}{n \cdot p \omega_{M}}\right]
    \left\{\mathcal{V}^{\rm{HT}}_{\text{NLP}}\left(q^{2}\right),
    \mathcal{A}^{\rm{HT}}_{1,\text{NLP}}\left(q^{2}\right),
    \mathcal{T}^{\rm{HT}}_{1,\text{NLP}}\left(q^{2}\right), \mathcal{T}_{2,\text{NLP}}^{\mathrm{HT}}\left(q^{2}\right)
    \right\} \\
    &=\frac{\tilde{f}_{B}(\mu) m_{B}}{(n \cdot p)^2}
    \left\{
    f_{2,1}[\boldsymbol{\tau}_1]+f_{3,2}[\boldsymbol{\tau}_2]-\kappa_i\frac{m_q}{n\cdot p}f_{3,2}[\boldsymbol{\tau}_2]
    \right\},\\
 f^\parallel_V & \exp\left[-\frac{m_{V}^{2}}{n \cdot p \omega_{M}}\right]
    \left\{\mathcal{A}^{\rm{HT}}_{0,\text{NLP}}\left(q^{2}\right),
    \mathcal{A}^{\rm{HT}}_{12,\text{NLP}}\left(q^{2}\right),
    \mathcal{T}^{\rm{HT}}_{23,\text{NLP}}\left(q^{2}\right)
    \right\} \\
    &=\frac{2\tilde{f}_{B}(\mu) m_{B}m_V}{(n \cdot p)^3}
    \left\{f_{2,1}[\boldsymbol{\tau}_1]
+f_{3,2}[\boldsymbol{\tau}_3]+\frac{m_q}{n \cdot p}f_{3,2}[\boldsymbol{\tau}_4]
+\iota_i \left(f_{3,2}[\boldsymbol{\tau}_5]+\frac{m_q}{n \cdot p}f_{3,2}[-\boldsymbol{\tau}_3]\right)\right\},
    \end{aligned}
\end{equation}
with the symmetry-breaking factors 
\begin{equation}
    \kappa_i\in\left\{
    +1,-1,\frac{n\cdot q}{\bar{n}\cdot q},-\frac{n\cdot q}{\bar{n}\cdot q}
    \right\},\quad\iota_i\in\left\{\frac{n\cdot q}{m_B},-\frac{n\cdot q}{m_B},-1\right\},
\end{equation}
where the function $f_{i,j}$ describes the contribution of terms in the form $\phi(\omega)/(\omega-\cdots)^j$, with $\phi(\omega)$ being the $i$-particle LCDA and the denominator raised to the $j$-th power. The explicit expressions of $f_{i,j}$ are listed in Appendix~\ref{dis func-app}, and the density functions are expressed by
\begin{equation}
    \begin{aligned}
     &\boldsymbol{\tau}_1(\omega)=4\frac{d}{d \omega}
     \hat{g}^{-}_{B}(\omega),
&\boldsymbol{\tau}_2(\omega_1,\omega_2,u)&=\Psi_5(\omega_1,\omega_2)+\tilde{\Psi}_5(\omega_1,\omega_2),
\\
    &\boldsymbol{\tau}_3(\omega_1,\omega_2,u)=
    \Psi_5(\omega_1,\omega_2)-\tilde{\Psi}_5(\omega_1,\omega_2),
    &\boldsymbol{\tau}_4(\omega_1,\omega_2,u)&=
    2\Phi_6(\omega_1,\omega_2),
\\
    &\boldsymbol{\tau}_5(\omega_1,\omega_2,u)=
    2\bar{u}\Phi_4(\omega_1,\omega_2). & &
    \end{aligned}
\end{equation}

\subsection{Higher-order terms in hard-collinear propagator}

Adopting the approach detailed in Refs.~\cite{Cui:2022zwm, Gao:2021sav, Cui:2023jiw}, we carry out the calculation of the subleading  power corrections stemming from the hard-collinear propagator in the correlation functions expressed by Eq.(\ref{cft}). Given that the momentum in the interpolating current lies in the hard-collinear regime, the quark propagator connecting this interpolating current and a soft quark also possesses hard-collinear momentum. Expanding the hard-collinear propagator in terms of powers of ${\rm{\Lambda}}_{\rm{QCD}}/m_b$ gives rise to
\begin{align}\label{HCP}
\frac{\slashed{p}-\slashed{k}+m_q}{(p-k)^2-m^{2}_{q}+i\epsilon}=\frac{\bigg[\overbrace{n\cdot p\frac{\slashed{\bar{n}}}{2}}^{\rm{LP}}+\overbrace{(\bar{n}\cdot p\frac{\slashed{{n}}}{2}-\slashed{k}+m_q)
 +\frac{n\cdot p\frac{\slashed{\bar{n}}}{2}\bar{n}\cdot p n\cdot k}{{n\cdot p(\bar{n}\cdot p-\bar{n}\cdot k)}}+\frac{n\cdot p\frac{\slashed{\bar{n}}}{2}(m^{2}_{q}-m^{2}_{q'})}{n\cdot p(\bar{n}\cdot p-\bar{n}\cdot k)}}^{\rm{NLP}}\bigg]}{n\cdot p(\bar{n}\cdot p-\bar{n}\cdot k)},
\end{align}
where $m_q$ is the mass of the hard-collinear  quark propagator and $m_{q'}$ is the mass of the $B$-meson spectator quark. ``LP'' refers to the leading-power contribution of hard-collinear propagator at tree level , while ``NLP'' represents the subleading-power contributions resulting from the expansion of the hard-collinear propagator.

We then insert the NLP terms into the correlation functions ${\rm{\Pi}}_{\parallel,\perp}^{(a)}$, and apply the HQET operator identities \cite{Braun:2017liq,Neubert:1993mb}
\begin{equation}\label{odHQET}
    \begin{aligned}
    &v_\rho\frac{\partial}{\partial x_\rho}\bar{q}(x)\Gamma[x,0] h_{v}(0)=v\cdot \partial\bar{q}(x)\Gamma[x,0] h_{v}(0)+i\int^{1}_{0}du\bar{u}\bar{q}(x)[x,ux]x^{\lambda}g_sG_{\lambda\rho}(ux)[ux,0]v^{\rho}\Gamma h_{v}(0),
\\
 &i v\cdot \partial\langle 0|\bar{q}(x)\Gamma[x,0] h_{v}(0)|\bar{B}_v\rangle=\bar{\rm{\Lambda}}\langle 0|\bar{q}(x)\Gamma[x,0] h_{v}(0)|\bar{B}_v\rangle,
\\
 &\frac{\partial}{\partial x_\rho}\bar{q}(x)\gamma_\rho\Gamma[x,0] h_{v}(0)=-i\int^{1}_{0}duu\bar{q}(x)[x,ux]x^{\lambda}g_sG_{\lambda\rho}(ux)[ux,0]\gamma^{\rho}\Gamma h_{v}(0)+im_{q'}\bar{q}(x)\Gamma[x,0] h_{v}(0).
    \end{aligned}
\end{equation}
Taking advantage of the three-body light-cone HQET matrix element up to twist-six accuracy in Eq.~(\ref{3pLCDA}), we are able to derive the results for the first NLP term presented in Eq.~(\ref{HCP}),
\begin{equation}
    \begin{aligned}
     {\rm{\Pi}}^{{\rm{I}},\text{QPE}}_{\text{NLP}}&=
     \left[\frac{\tilde{f}_B(\mu)m_{B}}{2n\cdot p}\right]
     \left\{\int^{\infty}_{0}d\omega\frac{(\omega-2\bar{\Lambda})\phi^{+}_{B}(\omega)}{\bar{n}\cdot p-\omega}+(m_{q'}-m_q)\int^{\infty}_{0}d\omega\frac{\phi^{-}_{B}(\omega)}{\bar{n}\cdot p-\omega}\right.\\
    &\qquad \qquad \left.-\int^{1}_{0}du\int^{\infty}_{0}d\omega_1\int^{\infty}_{0}d\omega_2\frac{2[u\Phi_4(\omega_1,\omega_2)+\Psi_4(\omega_1,\omega_2)]}{(\bar{n}\cdot p-\omega_1-u\omega_2)^2}\right\}, \\
    {\rm{\tilde{\Pi}}}^{{\rm{I}},\text{QPE}}_{\text{NLP}}&=
     \left[\frac{\tilde{f}_B(\mu)m_{B}}{2n\cdot p}\right]
     \left\{\int^{\infty}_{0}d\omega\frac{(\omega-2\bar{\Lambda})}{\bar{n}\cdot p-\omega}\phi^{-}_{B}(\omega)\right.\\
     &\left.\qquad\qquad-\int^{1}_{0}du \int^{\infty}_{0}d\omega_1\int^{\infty}_{0}d\omega_2\frac{2 \bar{u} \Psi_5(\omega_1,\omega_2)}{(\bar{n}\cdot p-\omega_1-u\omega_2)^2}
     \right\},
    \end{aligned}
\end{equation}
where the hadronic parameter $\bar{\rm{\Lambda}}$ characterizes the 
``effective mass" of the $B$-meson state in HQET. It can be defined as \cite{Falk:1992fm}
\begin{align}
 \bar{\rm{\Lambda}}\equiv\frac{\langle 0|\bar{q}iv\cdot \overleftarrow{D}\Gamma h_v|\bar{B}_q(v)\rangle
}{\langle 0|\bar{q}i\Gamma h_v|\bar{B}_q(v)\rangle}.
\end{align}
Noting the relation $n_\alpha=2v_\alpha-\bar{n}_\alpha$, we can further compute the contribution of the second NLP term along the same lines,
\begin{equation}
    \begin{aligned}
     {\rm{\Pi}}^{{\rm{II}},\text{QPE}}_{\text{NLP}}=0, \qquad
     {\rm{\tilde{\Pi}}}^{{\rm{II}},\text{QPE}}_{\text{NLP}}&=
        \left[\frac{\tilde{f}_B(\mu)m_{B}}{2n\cdot p}\right]
     \left\{\int^{\infty}_{0}d\omega\frac{\bar{n}\cdot p(2\bar{\Lambda}-\omega)}{(\bar{n}\cdot p-\omega)^2}\phi^{-}_{B}(\omega) \right.\\
     \qquad\qquad&+\left. \int^{1}_{0}du \int^{\infty}_{0}d\omega_1\int^{\infty}_{0}d\omega_2\frac{4 \bar{u} \, \bar{n}\cdot p \Psi_5(\omega_1,\omega_2)}{(\bar{n}\cdot p-\omega_1-u\omega_2)^3}\right\}.
    \end{aligned}
\end{equation}
 Subsequently, the contribution of the third NLP term of the hard-collinear propagator expansion can be easily derived by applying the standard factorization procedure,
\begin{equation}
    \begin{aligned}
     {\rm{\Pi}}^{{\rm{ III}},\text{QPE}}_{\text{NLP}}=0,\qquad
     {\rm{\tilde{\Pi}}}^{{\rm{III}},\text{QPE}}_{\text{NLP}}=
     (m_q^2-m_{q'}^2)\int^{\infty}_{0}d\omega\frac{1}{(\bar{n}\cdot p-\omega)^2}\phi^{-}_{B}(\omega).
    \end{aligned}
\end{equation}
Ultimately, we obtain the factorization formulas for the NLP contributions stemming from the quark propagator expansion at tree level:
\begin{equation}
\begin{aligned}
   {\rm{\Pi}}^{(a),\text{QPE}}_{j,\text{NLP}}=\Gamma^{(a)}_j( {\rm{\Pi}}^{{\rm{ I}},\text{QPE}}_{\text{NLP}}+ {\rm{\Pi}}^{{\rm{ II}},\text{QPE}}_{\text{NLP}}+ {\rm{\Pi}}^{{\rm{ III}},\text{QPE}}_{\text{NLP}})
   +\tilde{\Gamma}^{(a)}_j
   ({\rm{\tilde{\Pi}}}^{{\rm{I}},\text{QPE}}_{\text{NLP}}+
   {\rm{\tilde{\Pi}}}^{{\rm{II}},\text{QPE}}_{\text{NLP}}+
   {\rm{\tilde{\Pi}}}^{{\rm{III}},\text{QPE}}_{\text{NLP}}),
\end{aligned}
\end{equation}
with 
\begin{equation}
    \begin{aligned}
     & \Gamma_{\parallel,\perp}^{(\text{V}-\text{A})}=\{n_{\mu},g_{\delta\mu}^{\perp}-i\epsilon^\perp_{\delta\mu}\},\quad\quad \tilde{\Gamma}_{\parallel,\perp}^{(\text{V}-\text{A})}=\{\bar{n}_{\mu},g_{\delta\mu}^{\perp}+i\epsilon^\perp_{\delta\mu}\},
    \end{aligned}
\end{equation} and 
\begin{equation}
    \begin{aligned}
      & \Gamma_{\parallel,\perp}^{(\text{T}+\tilde{\rm{T}})}=-\left\{n_{\mu}\frac{\bar{n}\cdot q}{2}-\bar{n}_{\mu}\frac{n\cdot q}{2},(g_{\delta\mu}^{\perp}-i\epsilon^\perp_{\delta\mu})\frac{n\cdot q}{2}\right\},\\ & \tilde{\Gamma}_{\parallel,\perp}^{(\text{T}+\tilde{\rm{T}})}=-\left\{\bar{n}_{\mu}\frac{n\cdot q}{2}-n_\mu\frac{\bar{n}\cdot q}{2},(g_{\delta\mu}^{\perp}+i\epsilon^\perp_{\delta\mu})\frac{\bar{n}\cdot q}{2}\right\}.
    \end{aligned}
\end{equation}
Adopting the standard LCSR strategy, we formulate the correlation functions in the dispersion relation formalism and match them to the hadronic representation given from Eq.~(\ref{hcf}) to Eq.~(\ref{hcf1}). Finally, the desired $B\to K^*$ form factors for the power-suppressed contribution stemming  from hard-collinear propagator can be expressed as
\begin{equation}
    \begin{aligned}
    f_V \exp \left[ -\frac{m_V^2}{n\cdot p \, \omega_M} \right]
    \mathcal{F}_{i,\text{NLP}}^{\text{QPE}}(q^2)&=  \frac{\tilde{f}_B(\mu) m_B}{(n\cdot p)^2}
    \left\{\kappa_i\left( f_{2,1}[\boldsymbol{\eta}_1]-f_{3,2}[\boldsymbol{\eta}_2] \right) \right.\\
    \quad \quad &\left.+\tilde{\kappa}_i\left( f_{2,1}[\boldsymbol{\eta}_3]-f_{3,2}[\boldsymbol{\eta}_4]- f_{2,2}[\boldsymbol{\eta}_5]-f_{3,3}[\boldsymbol{\eta}_6] \right)\right\},
    \end{aligned}
\end{equation}
where form factors $\mathcal{F}_i\in\left\{ \mathcal{V},\mathcal{A}_1,\mathcal{T}_1,\mathcal{T}_2,\mathcal{A}_0,\mathcal{A}_{12},\mathcal{T}_{23}  \right\}$ with the replacement $f_V\to f_V^{\perp}$ for the first four $\mathcal{F}_i$ and $f_V\to f_V^{\parallel}$ for the last three $\mathcal{F}_i$, and the corresponding symmetry-breaking factors read
\begin{equation}
    \kappa_i\in\left\{
    1,-1,\frac{n\cdot q}{m_B},-\frac{n\cdot q}{m_B},-\frac{2m_V \, n\cdot q}{m_B \, n\cdot p},
\frac{2m_V \, n\cdot q}{m_B \, n\cdot p},\frac{2m_V}{n\cdot p}
    \right\},
\end{equation}

\begin{equation}
    \tilde{\kappa}_i\in\left\{
    1,1,\frac{\bar{n}\cdot q}{m_B},\frac{\bar{n}\cdot q}{m_B},\frac{2m_V }{ n\cdot p},
\frac{2m_V }{ n\cdot p},\frac{2m_V}{n\cdot p}
    \right\},
\end{equation}
and the density functions $\boldsymbol{\eta}_i$ are given as follows
\begin{equation}
    \begin{aligned}
     \boldsymbol{\eta}_1(\omega)&=(\omega-2\bar{\Lambda})\phi_B^+(\omega)
    +(m_{q'}\pm m_q)\phi_B^-(\omega), &
    \boldsymbol{\eta}_2(\omega_1,\omega_2,u)&=2[u\Phi_4(\omega_1,\omega_2)+\Psi_4(\omega_1,\omega_2)], \\
    \boldsymbol{\eta}_3(\omega)&=0, &
    \boldsymbol{\eta}_4(\omega_1,\omega_2,u)&=2 \bar{u} \Psi_5(\omega_1,\omega_2), \\
    \boldsymbol{\eta}_5(\omega)&=\omega(2\bar{\Lambda}-\omega)\phi_B^-(\omega)+(m_q^2-m_{q'}^2)\phi_B^-(\omega), &
    \boldsymbol{\eta}_6(\omega_1,\omega_2,u)&=2(\omega_1+u\omega_2)\boldsymbol{\eta}_4(\omega_1,\omega_2,u),
\end{aligned}
\end{equation}
where the $+$ sign and $-$ sign in $\boldsymbol{\eta}_1(\omega)$ assigned to  $\mathcal{F}_i\in\left\{ \mathcal{V},\mathcal{A}_1,\mathcal{T}_1,\mathcal{T}_2  \right\}$ and $\mathcal{F}_i\in\left\{ \mathcal{A}_0,\mathcal{A}_{12},\mathcal{T}_{23}  \right\}$, respectively.

\subsection{Subleading heavy-quark effective current}
We now proceed to consider  the contributions of the power-suppressed terms in the heavy quark expansion to the  $B \to K^*$ form factors. In HQET, the bottom quark can be replaced by the effective heavy quark field, and the heavy-to-light weak current is expanded up to NLP accuracy~\cite{Beneke:2002ni, Beneke:2018wjp},
\begin{align}
\bar{q}\Gamma_\mu b=\underbrace{\bar{q}\Gamma_\mu h_v}_{\rm{LP}}
+\underbrace{\frac{1}{2m_b}\bar{q}\Gamma_\mu 
i\overrightarrow{\slashed{D}}h_v}_{\rm{NLP}}+\cdots,
\end{align}
where~$\overrightarrow{\slashed{D}}=\slashed{D}-(v\cdot D)\slashed{v}$~and~$\overrightarrow{\slashed{D}}h_v(0)=[\slashed{D}-(v\cdot D)\slashed{v}]h_v(0)=\slashed{D}h_v(0)$ due to the HQET equations of motion.
The ellipses denote the terms in  powers of $\mathcal{O}(1/m^2_b)$, whose contributions to the correlation functions are beyond the scope of our current work.
Substituting the heavy-to-light effective current in the correlation functions with the NLP term and taking advantage of the operator identities in Eq.~(\ref{odHQET}) and the following equation,
\begin{equation}\label{OPI 2}
    \begin{aligned}
    \bar{q}(x)\Gamma[x,0]\overrightarrow{D_\rho} h_{v}(0)=&\partial_\rho[\bar{q}(x)\Gamma[x,0] h_{v}(0)]+i\int^{1}_{0}du\bar{u}\bar{q}(x)[x,ux]g_sG_{\lambda\rho}(ux)[ux,0]x^{\lambda}\Gamma h_{v}(0)
\\
 &-\frac{\partial}{\partial x^\rho}\bar{q}(x)\Gamma[x,0] h_{v}(0),
    \end{aligned}
\end{equation}
 the correlation functions    can be directly presented in the following form
\begin{equation}
\begin{aligned}
{\rm{\Pi}}^{(a),\text{HQE}}_{j,\text{NLP}}(p,q)&=
    \int d^4 xe^{ip\cdot x}\langle 0|T\{\bar{q}'(x) \Gamma_j q(x),\frac{1}{2m_b}\bar{q}(0)\Gamma_\mu^{(a)}i\overrightarrow{\slashed{D}}h_v\}|\bar{B}(p+q)\rangle
\\
&=\frac{-1}{2m_b}\int d^4 x\int 
        \frac{d^4 k}{(2\pi)^4}\frac{e^{ik\cdot x}}{\bar{n}\cdot p-\bar{n}\cdot k+i\epsilon}
\left\{\partial_\xi[\bar{q}(x)\Gamma_j\frac{\slashed{\bar{n}}}{2}\Gamma^{(a)}_\mu\gamma_\xi h_{v}(0)]\right.
\\
&\left.+i\int^{1}_{0}du\bar{u}\bar{q}(x)x^{\rho}g_sG_{\rho\xi}(ux)\Gamma_j\frac{\slashed{\bar{n}}}{2}\Gamma^{(a)}_\mu\gamma^{\xi}h_{v}(0)
-\frac{\partial}{\partial x_\xi}[\bar{q}(x)]\Gamma_j\frac{\slashed{\bar{n}}}{2}\Gamma^{(a)}_\mu\gamma_\xi h_{v}(0)\right\},
\end{aligned}    
\end{equation}
where $a\in\{\text{V}-\text{A},\text{T}+\tilde{\mathrm{T}}\}$ denotes the different Dirac structures $\Gamma_\mu^{(a)}\in$ $\{\gamma_{\mu}(1-\gamma_5),i\sigma_{\mu\nu}(1+\gamma_5)q^{\nu}\}$ and $\Gamma_j\in\{\frac{\slashed{n}}{2},\frac{\slashed{n}}{2}\gamma^{\perp}_{\delta}\}$ for $j=\parallel,\perp$, respectively.  
We can further derive the factorization formula by utilizing analogous techniques at tree level
\begin{equation}\label{HQE-2pt-fac}
\begin{aligned}
{\rm{\Pi}}_{j,\text{NLP}}^{(a),\text{HQE}}=&\Gamma_j^{(a)}\frac{\tilde{f}_B(\mu)m_B}{4m_b}\left\{
   \int_0^{\infty}d\omega\frac{1}{\bar{n}\cdot p-\omega}
    \left[ (2\bar{\Lambda}-\omega)\phi_B^+(\omega)+(\bar{\Lambda}-\omega-m_{q'})\phi_B^-(\omega)
      \right]      
\right.\\
&\left.\qquad+
\int_0^{\infty}d\omega_1\int_0^{\infty}d\omega_2\int_0^{1}du
\frac{2[\Phi_4(\omega_1,\omega_2)+\Psi_4(\omega_1,\omega_2)]}{(\bar{n}\cdot p-\omega_1-u\omega_2)^2}
     \right\},
\end{aligned}
\end{equation}
with
\begin{equation}
    \Gamma_{\parallel,\perp}^{\text{V}-\text{A}}\in\left\{-\bar{n}_\mu,
    g_{\delta\mu}^{\perp}+i\epsilon_{\delta\mu}^\perp
    \right\},\quad
    \Gamma_{\parallel,\perp}^{\text{T}+\tilde{\mathrm{T}}}\in\left\{n_\mu\frac{\bar{n}\cdot q}{2}-\bar{n}_\mu\frac{n\cdot q}{2},
    \bar{n}\cdot q(g_{\delta\mu}^{\perp}+i\epsilon_{\delta\mu}^\perp)
    \right\}.
\end{equation}
By matching the partonic representation with the hadronic dispersion relations given from Eq.~(\ref{hcf}) to Eq.~(\ref{hcf1}), we can derive the subleading-power heavy-quark effective current correction to the $B\to K^*$ form factors 
\begin{equation}
    \begin{aligned}
    f_V \exp \left[ -\frac{m_V^2}{n\cdot p \, \omega_M} \right]
    \mathcal{F}_{i,\text{NLP}}^{\text{HQE}}(q^2)&= \frac{\tilde{f}_B(\mu) m_B}{2(n\cdot p)m_b}
   c_i
   \left\{ f_{2,1}[\boldsymbol{\zeta}_1]+f_{3,2}[\boldsymbol{\zeta}_2]\right\},
    \end{aligned}
\end{equation}
where the coefficient factors $c_i$ are determined from the correlation functions at the hadronic level,
\begin{equation}
    c_i\in\left\{
    -1,-1,\frac{\bar{n}\cdot q}{m_B},\frac{\bar{n}\cdot q}{m_B},\frac{2m_V}{n\cdot p},\frac{2m_V}{n\cdot p},\frac{-2m_V}{n\cdot p}
    \right\},
\end{equation}
and the density functions $\boldsymbol{\zeta}_i$ can be derived from the correlation functions at the partonic level in Eq.~(\ref{HQE-2pt-fac})  
\begin{equation}
    \begin{aligned}
    \boldsymbol{\zeta}_1(\omega)=(2\bar{\Lambda}-\omega)\phi_B^+(\omega)+(\bar{\Lambda}-\omega-m_{q'})\phi_B^-(\omega),\,
    \boldsymbol{\zeta}_2(\omega_1,\omega_2,u)=2[\Phi_4(\omega_1,\omega_2)+\Psi_4(\omega_1,\omega_2)].
    \end{aligned}
\end{equation}
Both the correction from the hard-collinear propagator and the correction from the subleading effective current to the $B\to K^*$ form factors show excellent agreement with the previous calculation for the  $B\to D^*$ process reported in Ref.~\cite{Cui:2023jiw}, upon substituting the charm quark mass with the strange quark mass.

\subsection{Higher-twist four-particle contribution}
\begin{figure}
\centering
\includegraphics[width=0.75\textwidth]{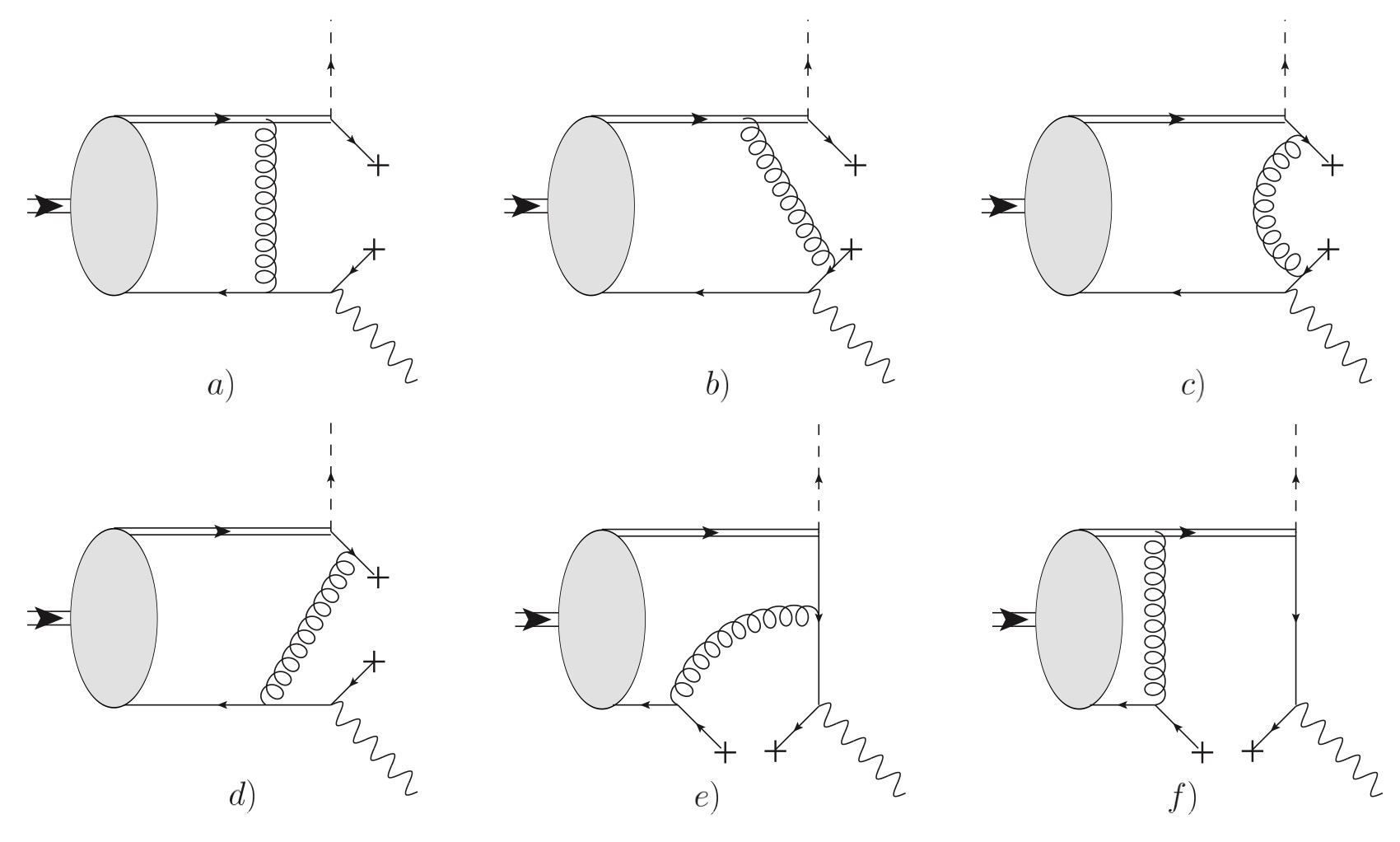}
\caption{Diagrammatic representations of twist-five and twist-six four-particle corrections to the vaccumm-to-bottom meson correlation functions.}
\label{fig:fourparticle}
\end{figure}
We are now in the position to calculate the heavy-to-light $B$-meson decay form factors from the twist-five and twist-six four-particle LCDAs in the factorization approximation. The subleading-power correction to the $B\to K^*$ form factors can be factorized into a product of the lower-twist two-particle LCDAs and the quark condensate~\cite{Beneke:2018wjp, Agaev:2010aq}. By evaluating the lowest-order Feynman
diagrams shown in Fig.~\ref{fig:fourparticle}, we obtain the non-leading Fock-state correction to the correlation functions
\begin{equation}
    \begin{aligned}
    {\rm{\Pi}}_{j,\text{NLP}}^{(a),4\text{P}}=&\frac{g^{2}_{s}C_F}{12}
     \frac{\tilde{f}_B(\mu)m_{B}}{n\cdot p}
\left\{\Gamma_j^{(a)}\langle \bar{q}q\rangle\int^{\infty}_{0}d\omega\frac{\phi^{+}_{B}(\omega)}{\bar{n}\cdot p(\omega-\bar{n}\cdot p)^2}
\right.
\\
&+\tilde{\Gamma}_j^{(a)}\langle \bar{q}'q'\rangle
\int^{\infty}_{0}d\omega
\frac{\phi^{+}_{B}(\omega)}{\omega^3}
\left.\bigg[\frac{2\omega}{\bar{n}\cdot p-\omega}-\frac{\omega^2}{\bar{n}\cdot p(\bar{n}\cdot p-\omega)}+2\ln{\frac{\bar{n}\cdot p-\omega}{\bar{n}\cdot p}}\bigg]\right\},
\end{aligned}
\end{equation}
with
$\Gamma_\parallel^{(a)}=\tilde{\Gamma}_\parallel^{(a)}=\left\{
\bar{n}_\mu,n_\mu\frac{\bar{n}\cdot q}{2}-\bar{n}_\mu\frac{n\cdot q}{2}\right\}$ and $\Gamma_\perp^{(a)}=0,$ $\tilde{\Gamma}_\perp^{(a)}=$ $\{
g_{\delta\mu}^{\perp}+i\epsilon_{\delta\mu}^\perp$,
$-\bar{n}\cdot q (g_{\delta\mu}^{\perp}+i\epsilon_{\delta\mu}^\perp)
\}$ for $a\in\{\text{V}-\text{A},\text{T}+\tilde{\mathrm{T}}\}$.  $\Gamma^{(a)}_j$ and $\tilde{\Gamma}_j^{(a)}$ denote the contributions from diagram-$(d)$ and diagram-$(e)$ in Fig.~\ref{fig:fourparticle}, respectively. The terms $\langle \bar{q}q\rangle$ and $\langle \bar{q}'q'\rangle$ represent the vacuum condensate of the propagator quark $q$ and the spectator quark $q'$, respectively.

It is worth mentioning that diagram~(e) in Fig.~\ref{fig:fourparticle} is analyzed using the background field expansion of the quark propagator on the light-cone \cite{Balitsky:1987bk}
\begin{align}
 \langle 0|T\{q(x),\bar{q}(0)\}|0\rangle\supset& \frac{\Gamma(d/2-1)}{8\pi^{d/2}(-x^2)^{d/2-1}}\int^{1}_{0}du u \bar{u}\slashed{x}g_sD_{\lambda}G^{\lambda\rho}(ux)x_\rho
 +\frac{\Gamma(d/2-2)}{16\pi^{d/2}(-x^2)^{d/2-2}}\nonumber\\
 &\int^{1}_{0}du (u \bar{u}-\frac{1}{2})g_sD_{\lambda}G^{\lambda\rho}(ux)\gamma_\rho,
\end{align}
where the classical equation of motion in QCD reads
\begin{align}
 D^{\lambda}G^{a}_{\lambda\rho}=-g_s\sum_{q}\bar{q}\gamma_{\rho}T^{a}q.
\end{align}
After explicitly calculating diagrams~(a),~(b) and~(c) in Fig.~\ref{fig:fourparticle}, the results indicate that, in comparison to the dominant contributions from diagrams (d) and (e), these three diagrams can only contribute to the higher power corrections. Moreover, diagram (f) is found to be insensitive to both the hard and hard-collinear QCD dynamics. Furthermore, diagram (d) is power-suppressed relative to diagram (e) when the interpolating current corresponds to a transversely polarized $K^*$.
 
By applying the dispersion relation to the correlation functions at the partonic level and employing the standard sum rule approach, we can derive the subleading-power corrections from four-particle contributions to the $B\to K^*$ form factors
 \begin{equation}
    \begin{aligned}
    f_V \exp \left[ -\frac{m_V^2}{n\cdot p \, \omega_M} \right]
    \mathcal{F}_{i,\text{NLP}}^{\text{4P}}(q^2)&= \frac{2\pi \alpha_s C_F \tilde{f}_B(\mu) m_B}{3(n\cdot p)^2}
   c_i\langle\bar{q}'q'\rangle
   \left\{
   \int_0^{\omega_s}d\omega \phi_{\text{eff-I}}^+(\omega)+
   \int_{\omega_s}^{\infty}d\omega\phi_{\text{eff-II}}^+(\omega)
   \right.\\
   &\,-\left.
   r_i\frac{\langle\bar{q}q\rangle}{\langle\bar{q}'q'\rangle}\left[ 
   e^{-\frac{\omega_s}{\omega_M}}\frac{\phi_B^+(\omega_s)}{\omega_s}-\int_{\omega_s}^{\infty}d\omega\frac{\phi_B^+(\omega)}{\omega^2}+
   \int_0^{\omega_s}d\omega \phi_{\text{eff-III}}^+(\omega)
   \right]
   \right\},
    \end{aligned}
\end{equation}
where the coefficients 
\begin{equation}
\begin{aligned}
 c_i\in\{1,1,\frac{\bar{n}\cdot q}{m_B},\frac{\bar{n}\cdot q}{m_B},\frac{2m_V}{n\cdot p},\frac{2m_V}{n\cdot p},\frac{2m_V}{n\cdot p}\},\,\,\,
    r_i\in\{0,0,0,0,1,1,1\},
\end{aligned}
\end{equation}
for $\mathcal{F}_i\in\left\{ \mathcal{V},\mathcal{A}_1,\mathcal{T}_1,\mathcal{T}_2,\mathcal{A}_0,\mathcal{A}_{12},\mathcal{T}_{23}  \right\}$ and the explicit expressions for the density functions $\phi_{\text{eff}}$ are given by
\begin{equation}
    \begin{aligned}
   \phi_{\text{eff-I}}^+(\omega)=&\left[1-2\frac{\omega_M}{\omega}+\left(1+2\frac{\omega_M}{\omega}\right)e^{-\frac{\omega}{\omega_M}}\right]\frac{\phi_B^+(\omega)}{\omega^2},\\
   \phi_{\text{eff-II}}^+(\omega)=&\left[1-2\frac{\omega_M}{\omega}+2\frac{\omega_M}{\omega}e^{-\frac{\omega_s}{\omega_M}}\right]\frac{\phi_B^+(\omega)}{\omega^2},\\
    \phi_{\text{eff-III}}^+(\omega)=&\left[\left(1+\frac{\omega}{\omega_M}\right)e^{-\frac{\omega}{\omega_M}}-1\right]\frac{\phi_B^+(\omega)}{\omega^2}.
    \end{aligned}
\end{equation}
Collecting the next-to-leading power (NLP) contributions estimated above, the total NLP correction to the $B\to K^*$ form factors in QCD can be expressed as
\begin{align}
F^{i}_{BK^*, \rm{NLP}}=F^{i,\rm{HT}}_{BK^*, \rm{NLP}}+
F^{i,\rm{QPE}}_{BK^*,\rm{NLP}}+F^{i,\rm{HQE}}_{BK^*,\rm{NLP}}
+F^{i,\rm{4P}}_{BK^*,\rm{NLP}},
\end{align}
where the index $i$ labels seven different $B\to K^*$ form factors in QCD. 
Before proceeding to the numerical analysis, we first establish the power-counting rules for the form factors at LP and NLP,
adopting the asymptotic behaviors of $B$-meson LCDAs~\cite{Braun:2017liq}. 
According to the power-counting scheme $\omega_M\sim\omega_s\sim\mathcal{O}(\lambda^2)$ and $m_s\sim\mathcal{O}(\lambda)$, where the scaling parameter $\lambda= {\rm{\Lambda}}_{\rm{QCD}}/m_b$, we derive the scaling behavior for the leading-power $B\to K^*$ form factors,
\begin{equation}
    \begin{aligned}
 &V_{BK^*}^{\rm{LP}}\sim A^{\rm{LP}}_{1,BK^*}\sim T^{\rm{LP}}_{1,BK^*}\sim T^{\rm{LP}}_{2,BK^*}\sim\mathcal{O}(\lambda^2),
\\
 &A^{\rm{LP}}_{0,BK^*}\sim\mathcal{O}(\lambda^3),\quad
{{A}_{12}}_{BK^*}^{\rm{LP}}\sim{{T}_{23}}_{BK^*}^{{\rm{LP}}}
\sim
\mathcal{O}(\lambda^4),
\end{aligned}
\end{equation}
where $ {A}_{12}$ and ${T}_{23}$ denote combinations of the form factors corresponding to $\frac{m_{B}+m_{V}}{n \cdot p}{A}_1-\frac{m_{B}-m_{V}}{m_{B}}{A}_2$ and $\frac{m_{B}}{n \cdot p}{T}_2-{T}_3$, respectively.
From the relations between seven QCD form factors and the four SCET effective form factors in Ref.~\cite{Gao:2019lta}, the scalings of the form factors ${V},{A}_1,{T}_1,{T}_2$ are determined by the SCET effective form factor
$\xi_\perp$, while ${A}_0, {A}_{12}, {T}_{23}$ are determined by the SCET effective form factor
$\xi_\parallel$. Evidently, $\xi_\parallel$ is suppressed by a factor of $\lambda^2$ compared to $\xi_\perp$. Because ${A}_0$ has
an enhancement factor $1/m_s$,
it ultimately contributes the power of $\mathcal{O}(\lambda^3)$. Applying the same analytical method as for the leading power and considering the asymptotic behavior of two-particle and three-particle higher-twist $B$-meson LCDAs,
we further determine the scalings of $B\to K^*$ form factors at NLP
\begin{equation}
    \begin{aligned}
 &V_{BK^*}^{\rm{NLP}}\sim A^{\rm{NLP}}_{1,BK^*}\sim T^{\rm{NLP}}_{1,BK^*}\sim T^{\rm{NLP}}_{2,BK^*}\sim\mathcal{O}(\lambda^3),
\\
 &A^{\rm{NLP}}_{0,BK^*}\sim\mathcal{O}(\lambda^4),\quad
{{A}_{12}}_{BK^*}^{\rm{NLP}}\sim{{T}_{23}}_{BK^*}^{{\rm{NLP}}}
\sim
\mathcal{O}(\lambda^5).
\end{aligned}
\end{equation}


\section{Numerical analysis}
Summing up both the NLL correction and the newly derived NLP corrections, we obtain the improved $B\to K^*$ form factors with SCET sum rules in the large recoil region.
In this section, we analyze phenomenological observables for the rare FCNC process $B\to K^* \nu_\ell\bar{\nu}_\ell$. 
We firstly list the relevant theoretical inputs in the factorization formula for the heavy-to-light form factors, including quark masses, decay constants, distribution amplitudes, sum rules parameters, electroweak parameters, and the inverse moments. We compare the form factors under two scenarios : selecting the inverse moment as either the recent lattice QCD simulation result $\lambda_B =389(35) $ MeV or the conventional range $\lambda_B =350(150) $ MeV. The theoretical precision of form factors could be improved if lattice QCD simulations systematically account for potential uncertainties \cite{Han:2024yun}.
After taking into account both the available lattice QCD results in the high-$q^2$ region and improved LCSR predictions in the low-$q^2$ region, we then perform a combined fit to determine the coefficients in the $z$-series expansion, thereby extending the $B \to K^*$ form factors to the entire kinematic region. 
Taking advantage of the newly updated LCSR predictions, we investigate the differential branching ratio and longitudinal $K^
*$ polarization fraction of $B\to K^{*} \nu_\ell \bar{\nu}_\ell$ decays.
For $B^+\to K^{*+} \nu_\ell \bar{\nu}_\ell$ decay process, the additional long-distance effect induced by $B^+\to\tau^+(\to K^{*+}\nu_{\tau})\bar{\nu}_{\tau}$ at tree level is included with the narrow $\tau$ width approximation.

\subsection{Theory inputs}
In Tab.~\ref{tab:para}, we summarize the necessary input parameters of the Standard Model and relevant hadronic parameters, along with the central values and uncertainties. 
In our numerical calculations, we employ the three-loop evolution of the strong coupling constant $\alpha_s(\mu)$ in the $\overline{\rm{MS}}$ scheme by taking the interval $\alpha^{(5)}_s(m_Z)$ from~\cite{ParticleDataGroup:2024cfk} and adopting the perturbative matching scales $\mu_4=4.8$ \rm{GeV} and $\mu_3=1.3$ \rm{GeV} for crossing the $n_f= 4$ and $n_f= 3$ thresholds, respectively \cite{Shen:2020hfq,Beneke:2020fot}. 
Additionally, the bottom quark mass $m_b (m_b)$ and strange quark mass $m_s (m_s)$ are given in the $\overline{\rm{MS}}$ scheme at the scale of their respective $\overline{\rm{MS}}$ masses. Using {\bf RunDec}~\cite{Chetyrkin:2000yt}, we obtain the scale dependence of the strong coupling constant $\alpha_s(\mu)$ and the quark masses $m_b(\mu)$ and $m_s(\mu)$. 
Moreover, we incorporate the results from four-flavor lattice QCD computations for the $B$-meson decay constant 
$f_B$~\cite{FlavourLatticeAveragingGroupFLAG:2021npn}. The decay constant of the longitudinal $K^*$ can be extracted from leptonic decays $V^0\to e^+e^-$ and tau lepton decays $\tau^+\to V^+\nu_\tau$, while the renormalization scale-dependent decay constant of the transverse $K^*$ is taken  from a lattice QCD simulation with $2+1$ flavors of domain wall quarks and the Iwasaki gauge action \cite{RBC-UKQCD:2008mhs}. Two hard scales $\mu_{h1}$ and $\mu_{h2}$ are introduced in the hard functions and $B$-meson decay constants, respectively. 
The factorization scale
$\mu$ is same as the hard-collinear scale and the renormalization scale for the QCD tensor current will be taken as $\nu=m_b$.
The LCSR improved form factors for the $B$-meson semileptonic decay processes depend on the $B$-meson light-cone distribution amplitudes (LCDAs) as universal non-perturbative input parameters.
Therefore, we need to construct an acceptable phenomenological model for the leading- and higher-twist $B$-meson LCDAs that not only satisfies the classical equations of motion~\cite{Braun:2017liq}, but also exhibits the expected asymptotic behavior at sufficiently large scales. 
In this work, we adopt a newly proposed three-parameter model for all the relevant $B$-meson light-cone distribution amplitudes in coordinate space \cite{Beneke:2018wjp}, with the details provided in Appendix~\ref{B meson LCDAs}. 
The three shape parameters $\alpha$, $\beta$ and $\omega_0$ in this model can be related to the inverse logarithmic moments $\lambda_B$ and $\hat{\sigma}_{1,2}$ for the leading-twist $B$-meson distribution amplitude $\phi^+_B$ with the equations
\begin{align}
 \lambda_B(\mu) &= \frac{\alpha-1}{\beta-1}\,\omega_0\,,
\nonumber \\
\widehat \sigma_1(\mu) &= 
\psi(\beta-1) -\psi(\alpha-1) + \ln\frac{\alpha-1}{\beta-1}\,,
\nonumber \\
\widehat \sigma_2(\mu)&= 
\widehat \sigma_1^2(\mu)+\psi'(\alpha-1)-\psi'(\beta-1)
+\frac{\pi^2}{6},
\end{align}
and the definitions of the inverse logarithmic moments $\lambda_B$ and $\hat{\sigma}_{1,2}$ are
\begin{align}
 \frac{1}{\lambda_B(\mu)} &=  
\int^\infty_0 \frac{d\omega}{\omega}\,\phi_B^+(\omega,\mu) \,,
\nonumber \\
\frac{\widehat{\sigma}_n(\mu)}{\lambda_B(\mu)} &= 
\int^\infty_0 \frac{d\omega}{\omega}\,
\ln^n\frac{e^{-\gamma_E}\lambda_B(\mu)}{\omega}\,
\phi_B^+(\omega,\mu) \, .
\end{align}
The numerical values for the hadronic parameters $\lambda_B$, $\hat{\sigma}_{1,2}$ and $\lambda_{E,H}$ in Tab.~\ref{tab:para} are all given at the reference scale $\mu_0=1$~GeV and these parameters will be evolved to the factorization scale $\mu$ in the final results.
Despite various strategies being employed to investigate the inverse moment $\lambda_B$ \cite{Beneke:2018wjp, Han:2024yun, Janowski:2021yvz, Ball:2003fq,Mandal:2023lhp,Lee:2005gza,Feldmann:2014ika,Braun:2003wx, Wang:2019msf}, a QCD-based method for its precise determination remains elusive due to its definition via a non-local operator (see Refs.~\cite{Han:2024yun,Han:2024min} for preliminary results from the lattice QCD perspectives).
We adopt a conservative interval of $\lambda_B=(350\pm150)$ MeV in this work and compare the resulting form factors with those derived from $\lambda_B=389(35)$ MeV, as suggested by the recent lattice QCD result \cite{Han:2024yun,Han:2024min}.
For the inverse logarithmic moments $\hat{\sigma}_{1,2}$, we prefer the choice $\{\hat{\sigma}_{1},\hat{\sigma}_{2}\}=\{0,\pi^2/6\}$ with the intervals
\begin{equation}
    \begin{aligned}
        -0.7<\hat{\sigma}_{1}<0.7,\qquad -6.0<\hat{\sigma}_{2}<6.0.
    \end{aligned}
\end{equation}
Following the standard procedure outlined in Refs.~\cite{Khodjamirian:2006st, Gao:2019lta, Ball:1998sk}, the two intrinsic parameters $\omega_M$ and $\omega_s$ introduced by the light-cone sum rules can be determined by effectively constraining the smallness of the continuum contributions in the dispersion integrals and the stability of the obtained sum rules results against the variation of $\omega_M$. The 
parameters $s_0^{\perp}$ and $s_0^{\parallel}$ correspond to the interpolating currents $\bar{q}'\slashed{n}\gamma_{\perp}q$ and $\bar{q}'\slashed{n}q$, respectively, leading to the following intervals
\begin{align}
&s_0^{\perp}=n\cdot p \,\omega^{\perp}_s=(1.4\pm 0.1)\,{\rm{GeV^2}},
\qquad
s_0^{\parallel}=n\cdot p \,\omega_s^{\parallel}=(1.7\pm 0.1)\,{\rm{GeV^2}},
\nonumber\\
&M^2=n\cdot p \,\omega_M=(1.7\pm 0.5)\,{\rm{GeV^2}}.
\end{align}

\begin{table}[htb] 
\centering \setlength\tabcolsep{4.5pt} \def\arraystretch{1.5}
\caption{Numerical values of the input parameters.}
\vspace{3pt}
\begin{tabular}{| l  l  l  | l  l  l |} 
\hline 
    Parameter & Value & Ref. & Parameter & Value & Ref.  \\ 
\hline 
\hline
$m_{B^0}$ & $ 5279.66  $ MeV & \cite{ParticleDataGroup:2024cfk} & 
$m_{{K^{*}}^0}$ & $898.46$ MeV & \cite{ParticleDataGroup:2024cfk}\\
\hline 
$m_{B^+}$ & $5279.34  $ MeV & \cite{ParticleDataGroup:2024cfk} & 
$m_{{K^{*}}^+}$ & $ 891.67 $ MeV & \cite{ParticleDataGroup:2024cfk}\\
\hline 
$\tau_{B^0}$ & 1.517(4)\,ps  & \cite{ParticleDataGroup:2024cfk} & $\tau_{B^+}$
 & 1.638(4) \,ps & \cite{ParticleDataGroup:2024cfk} \\
 \hline
$m_{\tau^+}$ & $1776.86 $ MeV & \cite{ParticleDataGroup:2024cfk} &
$\tau_{\tau^+}$ & $0.2903(5)$ ps & \cite{ParticleDataGroup:2024cfk}
\\
\hline 
$G_F$ & $1.166379\times10^{-5}\,\rm{GeV}^{-2}$  & \cite{ParticleDataGroup:2024cfk} & $\sin^2\theta_W$
 & 0.23126(5) & \cite{Brod:2010hi} \\
\hline 
$|V_{tb}V^{*}_{ts}|$ & $(41.25 \pm 0.45)\times10^{-3} $ & \cite{UTfit:2022hsi} & $\alpha_{em} (m_Z)$
 & ${1}/{127.925}$ & \cite{ParticleDataGroup:2024cfk} \\
 \hline
$|V_{ub}|$ & $3.82(20)\times 10^{-3} $  & \cite{ParticleDataGroup:2024cfk} &
$|V_{us}|$ & $0.2243(8) $ & \cite{ParticleDataGroup:2024cfk}
\\
\hline 
${m}_b({m}_b)$ & $(4.203\pm0.011)$ GeV & \cite{ParticleDataGroup:2024cfk} & 
${m}_s({m}_s)$ & $(93.5\pm 0.8)$ MeV & \cite{ParticleDataGroup:2024cfk} \\
\hline 
$f_B$ & $(190.0 \pm 1.3)$ MeV &\cite{FlavourLatticeAveragingGroupFLAG:2021npn} & 
$\mu_{h1}$& $[{m}_b/2,\, 2\,{m}_b]$  &  \\
\hline 
$f^{\parallel}_{K^*}$ & $(204\pm7)$ MeV &  \cite{Bharucha:2015bzk}
& 
$\mu_{h2}$ & $[{m}_b/2,\, 2\,{m}_b]$ &  \\
\hline 
$f^{\perp}_{K^*} (1 \rm{Gev})$ & $(159\pm6)$ MeV&\cite{RBC-UKQCD:2008mhs} &  
$\mu$ & $1.5\pm 0.5$ GeV &  \\
\hline 
 $\langle\bar{q}q\rangle  (2\rm{Gev})$ &$-(286\pm 23 \,\rm{MeV})^3$  & \cite{FlavourLatticeAveragingGroupFLAG:2021npn} & 
 $\nu$ & $m_b$ &  \\
 \hline 
 $\langle\bar{s}s\rangle:\langle\bar{q}q\rangle$& $(0.8\pm 0.1)$  & \cite{Ioffe:2002ee,Gelhausen:2013wia} & 
$M^2$ & $(1.7 \pm 0.5)$ GeV$^2$ & \cite{Ball:1998sk} \\
 \hline 
$s_{0}^{\parallel}$ & $(1.7 \pm 0.1)$ GeV$^2$ & \cite{Khodjamirian:2006st, Ball:1998sk} & 
$s_0^{\perp}$ & $(1.4 \pm 0.1)$ GeV$^2$ & \cite{Khodjamirian:2006st, Ball:1998sk} \\
\hline 
$\lambda_B$ & $350\pm150$ MeV&  \cite{Beneke:2018wjp}
& \multirow{3}{*}{$\{\widehat{\sigma}_1,\widehat{\sigma}_2\}$} 
& $\{0.7\,,6.0\}$ 
& \multirow{3}{*}{\cite{Beneke:2018wjp}} \\
\cline{1-3} 
$(\lambda^2_E/\lambda^2_H)$ & $0.50\pm0.10$ & \cite{Beneke:2018wjp}
&~  & $\{0.0\,,\pi^2/6\}$  &~  \\
\cline{1-3} 
$(2\lambda^2_E+\lambda^2_H)$ & $(0.25\pm0.15)$ GeV$^2$ 
&  \cite{Beneke:2018wjp}
&~  &$\{-0.7\,,-6.0\}$   &~ \\
\hline 
\end{tabular} 
\label{tab:para}
\end{table}
\begin{table}[htb]
\centering \setlength\tabcolsep{6.0pt} \def\arraystretch{1.5}
\caption{ The $B \to K^*$ form factors at $q^2=0$ given by our work (second row) and by sum rules with light-meson distribution amplitudes (third row).
}
\vspace{3pt}
\begin{tabular}{|c|c|c|c|c|c|c|c|}
\hline
 $\mathcal{F}_{i}^{B\to K^*}$&$\mathcal{V}$  &$\mathcal{A}_0$  &$\mathcal{A}_1$  &$\mathcal{T}_1$  &$\mathcal{T}_2$  & $\mathcal{A}_{12}$ & $\mathcal{T}_{23}$ \\ \hline
  This work& 0.20(14) & 0.066(38) & 0.19(13) & 0.24(16) & 0.19(13) & 0.066(38) & 0.094(44) \\ \hline
 Ref.~\cite{Bharucha:2015bzk}& 0.29(3) & 0.118(16) &0.306(33)  & 0.282(31)  &0.274(31)  &0.113(15)  &0.095(13)  \\ \hline
\end{tabular}

\label{tab: LCSRFFs}
\end{table}

\subsection{Numerical predictions for the \texorpdfstring{$B\to K^*$}{} form factors}
\begin{figure}[thb]
\begin{center}
\includegraphics[width=\columnwidth]{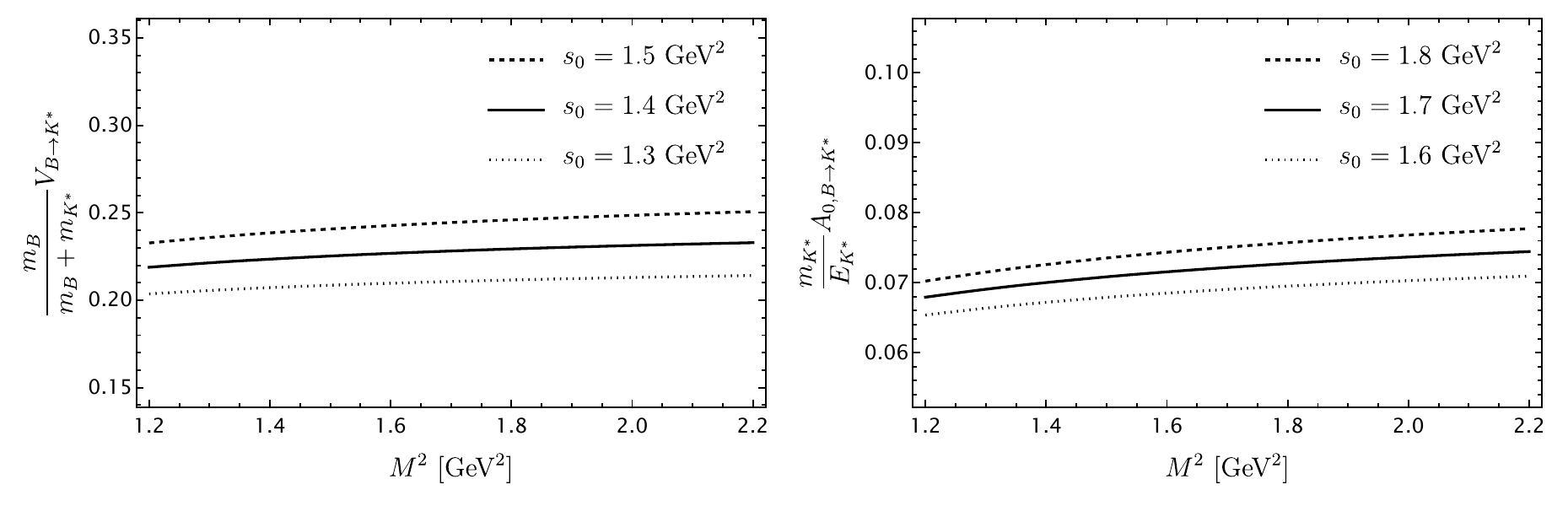}
\end{center}
\vspace{-20pt}
\caption{Dependence of the form factors $\mathcal{V}_{B\to K^{*}}$ and $\mathcal{A}_{0,B \to K^{*}}$ on the Borel parameter $M^2$.}
\label{fig:M2}
\end{figure}

Making use of the numerical inputs from Tab.~\ref{tab:para} and the $B$-meson light-cone distribution amplitudes described by three-parameter model in Appendix~\ref{B meson LCDAs}, we obtain the $B\to K^*$ form factors in the large recoil region. 
In Tab.~\ref{tab: LCSRFFs}, we present the numerical results of $B\to K^*$ form factors based on LCSR with heavy-meson distribution amplitudes at $q^2=0$. The central values of our improved form factors are consistent within a 1$\sim$2$\sigma$ deviation with the results obtained from sum rules based on $K^*$ distribution amplitudes~\cite{Bharucha:2015bzk}. 
To examine the numerical features of the LCSR parameters to form factors, we first present the dependence of the form factors on the Borel mass $M^2$ in Fig.~\ref{fig:M2}. 
The left panel in Fig.~\ref{fig:M2} shows the variation of form factor $\mathcal{V}$ when the Borel mass $M^2$ changes in the range of 1.2 $\rm{GeV}^2$ to 2.2 $\rm{GeV}^2$, with the effective threshold $s_0^{\perp}$ fixed at 1.3~$\rm{GeV}^2$, 1.4~$\rm{GeV}^2$, and 1.5~$\rm{GeV}^2$. The right panel in Fig.~\ref{fig:M2} illustrates the effect on the form factor $\mathcal{A}_0$ for the same Borel mass range, with the effective threshold set at 1.6~$\rm{GeV}^2$, 1.7~$\rm{GeV}^2$, and 1.8~$\rm{GeV}^2$, respectively. 
We find that LCSR form factors exhibit a mild dependence on the intrinsic parameters $M^2$ and $s_0^{\parallel, \perp}$, with each introducing a 10\% systematic uncertainties to the form factors, which is consistent with our previous work~\cite{Gao:2019lta} and other sum rules analyses~\cite{Cui:2022zwm,Gao:2021sav}.

\begin{figure}[thb]
\begin{center}
\includegraphics[width=\columnwidth]{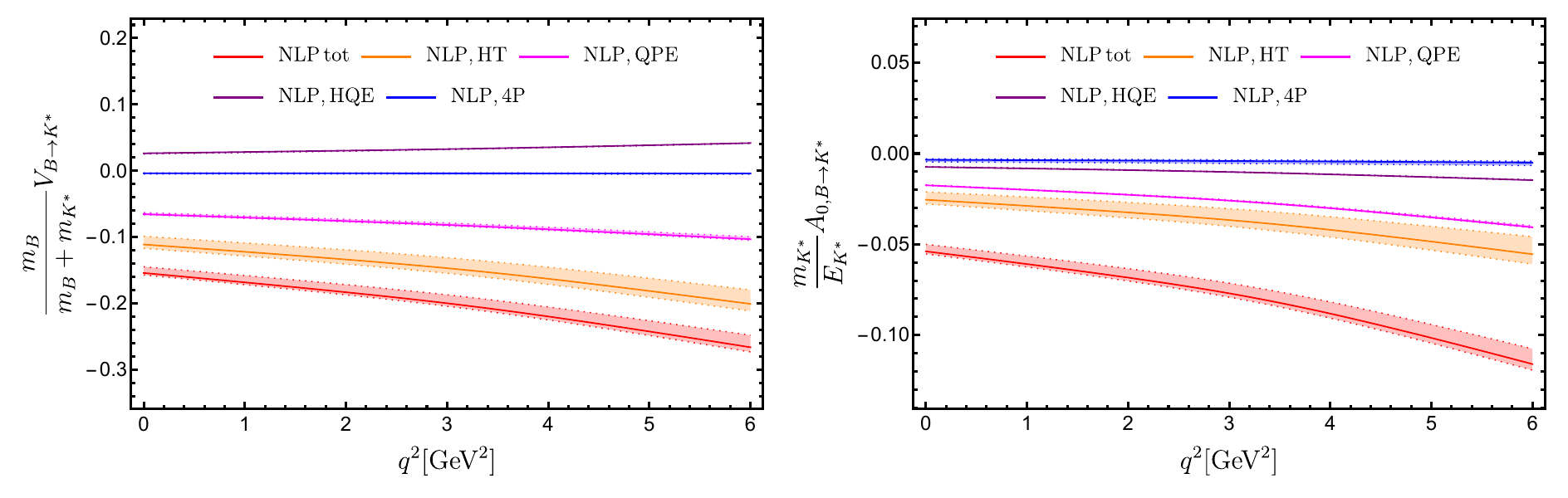}
\end{center}
\vspace{-0.5cm}
\caption{Subleading-power corrections to the $B\to K^{*}$ form factors $\mathcal{V}_{B\to K^{*}}$ (left panel) and $\mathcal{A}_{0,B \to K^{*}}$ (right panel) in the kinematic region of $0\leq q^2\leq 6\, \rm{GeV}^2$. The shaded bands represent the uncertainties from the variation of factorization scale $\mu$. }
\label{fig:q2SR}
\end{figure}

We now proceed to explore the contributions of the subleading-power corrections from different sources to the $B\to K^{*}$ form factors, with the form factors 
$\mathcal{V}$ and $\mathcal{A}_0 $ as illustrative examples. In Fig.~\ref{fig:q2SR}, we present the contributions of four different sources of the subleading-power corrections as well as the total power correction in the kinematic region of $0\leq q^2\leq 6 \, \rm{GeV}^2$. 
These subleading-power corrections include the `` HT " contribution from the two-particle and three-particle higher-twist  $B$-meson distribution amplitudes, the `` QPE " contribution from the expansion of hard-collinear quark propagator in the small parameter ${{\rm{\Lambda}}}_{\rm{QCD}}/m_b$, the `` HQE " contribution from the power-suppressed effective weak transition current $\bar{q}\Gamma[i\slashed{D}_{\perp}/(2m_b)]h_v$, and the `` 4P " contribution from twist-5 and twist-6 four-particle $B$-meson LCDAs within the factorization approach.
We can find that the NLP contribution from twist-5 and twist-6 four-particle $B$-meson LCDAs is minimal for the $B\to K^{*}$ form factors and this contribution accounts for only 3\% $\sim$ 7\% of the total NLP contribution to the form factors $\mathcal{V}$ and $ \mathcal{A}_0 $, respectively. In contrast, it is evident that the higher-twist $B$-meson LCDAs provide the largest contributions to the NLP $B\to K^{*}$ form factors, which numerically account for 50\% $\sim$ 60\% of the total NLP corrections in analogy to the previous discussions~\cite{Gubernari:2018wyi,Cui:2022zwm,Gao:2021sav, Wang:2018wfj}. 
\begin{figure}[thb]
\begin{center}
\includegraphics[width=\columnwidth]{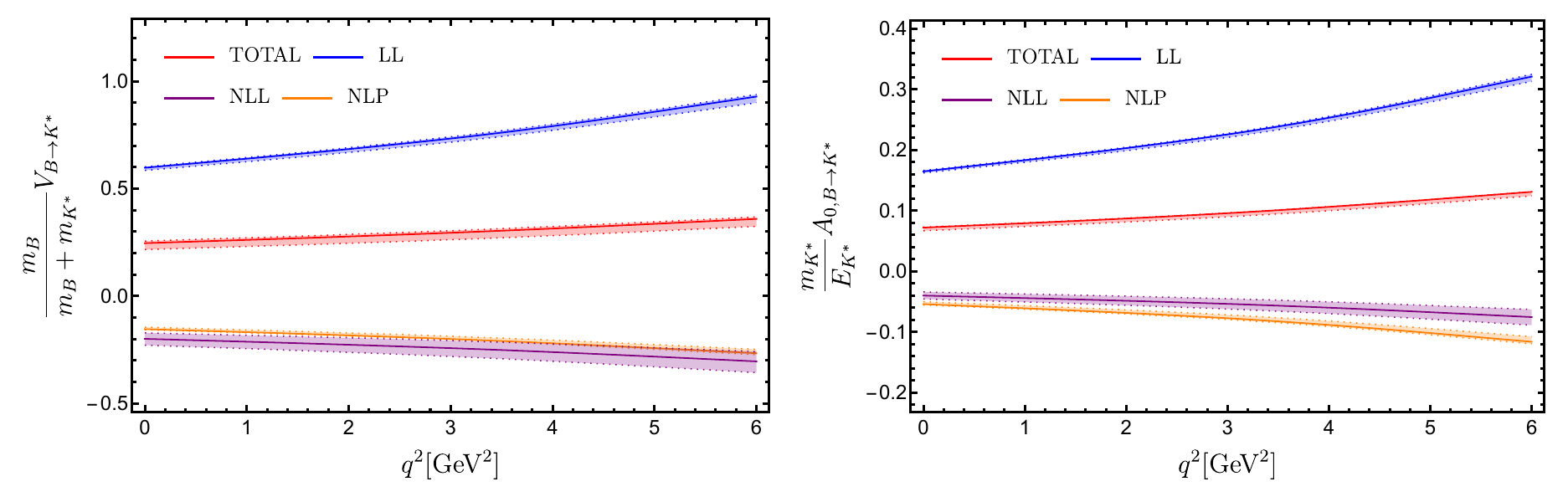}
\end{center}
\vspace{-0.5cm}
\caption{ Comparison of the LL resummation improved tree-level contribution (LL), NLL resummation improved one-loop correction (NLL), subleading-power correction at tree level (NLP), and total result (TOTAL) to the $B\to K^{*}$ form factors $\mathcal{V}_{B\to K^{*}}$ (left panel) and $\mathcal{A}_{0,B \to K^{*}}$ (right panel) in the kinematic region of $0\leq q^2\leq 6~\rm{GeV}^2$. The shaded bands represent the uncertainties from the variation of factorization scale $\mu$. }
\label{fig:q2tot}
\end{figure}

We now explore the contribution of the NLL resummation improved leading-power $B\to K^{*}$ form factors and the newly derived subleading-power corrections to the $B\to K^{*}$ form factors at tree level. 
In order to understand the impact of one-loop and subleading-power corrections, we show the numerical results explicitly for the resummation improved contribution at one-loop level and NLP corrections at the tree level to the $B\to K^{*}$ form factors in the region of $0\leq q^2\leq 6 \rm{GeV}^2$ in Fig.~\ref{fig:q2tot}.
For instance, the resummation improved NLL correction reduces the form factors $\mathcal{V}$ and $\mathcal{A}_0$ by 30\% compared to the results at leading-logarithm accuracy. 
In addition, the newly determined NLP corrections lead to an approximate 30$\%$ reduction to the form factors $\mathcal{V}$ and $\mathcal{A}_0$, respectively. 
After including both NLP and NLL corrections, we find that the total results for the form factors $\mathcal{V}$ and $\mathcal{A}_0$ exhibit a 60\% reduction relative to the corresponding LL resummation improved tree-level predictions.

\begin{figure}[thb]
\begin{center}
\includegraphics[width=\columnwidth]{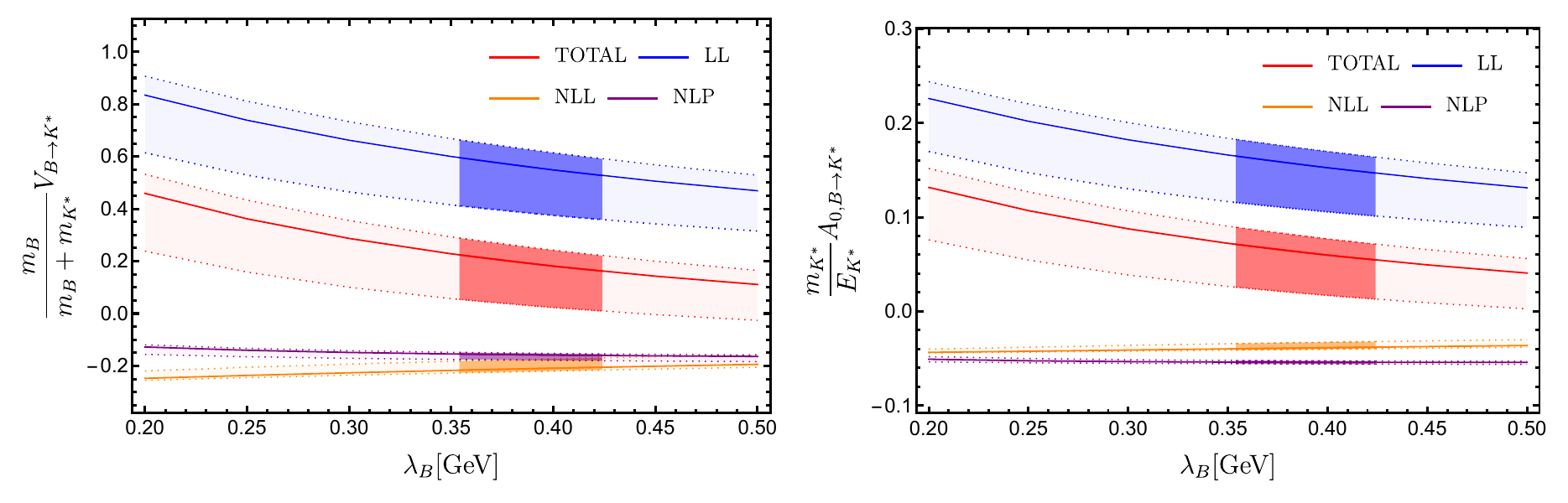}
\end{center}
\vspace{-0.5cm}
\caption{Comparison of LL resummation improved tree level contribution (LL),  NLL resummation improved one-loop correction (NLL), subleading-power correction at tree level (NLP) and total result (TOTAL) to the $B\to K^{*}$ form factors  $\mathcal{V}_{B\to K^{*}}$ (left panel) and $\mathcal{A}_{0,B \to K^{*}}$ (right panel) with the variation of $0.2\,\rm{GeV}\leq\lambda_B\leq 0.5 \,\rm{GeV}$. The areas with deeper color correspond to $\lambda_B=0.389(35)\,\rm{GeV}$ given in Ref. \cite{Han:2024min}. The upper (lower) bound represents $\{\hat{\sigma}_{1},\hat{\sigma}_{2}\}=\{-0.7,-6.0\}$ ($\{\hat{\sigma}_{1},\hat{\sigma}_{2}\}=\{0.7,6.0\}$). }
\label{fig:Lamtot}
\end{figure}

In addition, we investigate the dependence of both the NLL resummation improved one-loop correction and the newly derived NLP corrections on the inverse moment $\lambda_B$ at $q^2=0$. We take $\{\hat{\sigma}_{1},\hat{\sigma}_{2}\}=\{0,\pi^2/6\}$ as the central values and display the corresponding numerical results in Fig.~\ref{fig:Lamtot}.
For instance,
we observe that the form factors $\mathcal{V}_{B\to K^{*}}$ and $\mathcal{A}_{0,B \to K^{*}}$ exhibit a pronounced decrease with increasing  $\lambda_B $. By adopting $\lambda_B = 389(35)$ MeV as input  from lattice QCD in Ref. \cite{Han:2024min}, we find that both the central values and uncertainties of the form factors are reduced by approximately 20$\%$ compared to those obtained with $\lambda_B = 350(150)$ MeV. Notably, the light-cone sum rules based on light-meson LCDA suggest $\lambda_B \approx$ 300 MeV, when fitting $\lambda_B$ by using the form factor values from Ref.~\cite{Bharucha:2015bzk}.
This discrepancy between $\lambda_B \approx$ 300 MeV and the $\lambda_B$ = 389 MeV derived by lattice QCD may stem from unaccounted power corrections in current lattice QCD simulations, which could potentially introduce additional systematic uncertainties.

Since LCSR predictions are valid only in the large recoil region, it is necessary to extrapolate the LCSR results for the $B\to K^*$ form factors to the entire kinematic region by employing the BCL $z$-series expansion \cite{Okubo:1971my, Bourrely:1980gp, Bourrely:2008za, Lellouch:1995yv, Bourrely:2005hp}, which is based on the positivity and analyticity of the transition form factors. For this purpose, we apply the conformal transformation
\begin{align}
 z(q^2,t_0)=\frac{\sqrt{t_{+}-q^2}-\sqrt{t_{+}-t_0}}{\sqrt{t_{+}-q^2}+\sqrt{t_{+}-t_0}}
\end{align}
with the threshold parameter $t_{+}\equiv (m_B+m_{K^{*}})^2$ for the exclusive $B\to K^*$ form factors, which allows us to map the complex cut
$q^2$-plane onto the unit disk $|z(q^2,t_0)|\leq 1$. Additionally, the free parameter $t_0 < t_{+}$ corresponds to the value of $q^2$ that is mapped onto the origin in the $z$-plane. To minimize the 
$z$-interval, we set
\begin{align}
 t_0=t_{+}-\sqrt{t_{+}(t_{+}-t_{-})},\qquad
 t_{-}=(m_B-m_{K^{*}})^2.
\end{align}
Taking into account the asymptotic behavior of the form factor near the threshold of the corresponding excited states, we can further parameterize the $B\to K^*$ form factors with the $z$-series expansion as follows
\begin{align}\label{BCL}
 \mathcal{F}^{i}_{B\to K^*}(q^2)=\frac{1}{1-q^2/m^{2}_{i,\text{pole}}}\sum^{N-1}_{k=0} b^{i}_{k} \left[z(q^2,t_0)-z(0,t_0)\right]^{k},
\end{align}
where $m^{2}_{i,\text{pole}}$ denotes the masses of the corresponding resonances below the particle-pair production threshold $(m_B+m_{K^{*}})$ with distinct quantum numbers. 
For convenience, we have summarized the masses of the resonances relevant to our parameterization in Tab.~\ref{tab:pole} as given in Ref.~\cite{Bharucha:2015bzk}. 
We will truncate the $z$-series expansion at $N=3$ in the subsequent fitting process, since the contribution beyond quadratic terms is negligible due to $|z(q^2)|_{\text{max}}<0.1$.

\begin{table}[thb] 
\centering \setlength\tabcolsep{8pt} \def\arraystretch{1.5}
\caption{Summary of the resonance masses with distinct quantum numbers appearing in the $z$-series expansion of the $B\to K^*$ form factors in 
Eq.~\eqref{BCL}.}
\vspace{4pt}
\begin{tabular}{| l  c  c |} 
\hline 
$\mathcal{F}^{i}_{B\to K^*}(q^2)$ & $J^P$ & $m_{i,pole}$ [GeV]\\
\hline 
$\mathcal{V}(q^2)$, $\mathcal{T}_1(q^2)$ & $1^-$ & 
5.415 \\
\hline 
$\mathcal{A}_0(q^2)$ & $0^-$ & 
5.366 \\
\hline 
$\mathcal{A}_1(q^2)$, $\mathcal{A}_{12}(q^2)$, $\mathcal{T}_2(q^2)$, $\mathcal{T}_{23}(q^2)$ & $1^+$ & 
5.829 \\
\hline 
\end{tabular} 
\label{tab:pole}
\end{table}

We are now in the position to determine the $z$-series coefficients $b_{0,1,2}^i$ of the $B\to K^*$ form factors $\mathcal{F}^i(q^2)$ by performing the correlated minimum-$\chi^2$ fit of 
the updated LCSR predictions in the large recoil region, in combination with the available lattice QCD data in the small recoil region \cite{Horgan:2013hoa, Horgan:2015vla}. The ingredients of the minimum-$\chi^2$ fit can be summarized as follows:
\begin{itemize}
    \item In the low $q^2$ region, we generate the improved LCSR form factors with uncertainties at three distinct kinematic points $q^2=\{-4, 0, 4\}\,{\rm{GeV}}^2$. In order to obtain the pseudo-data samples, we vary the theoretical input parameters randomly within the error ranges and generate an ensemble of $N=300$ parameter sets that follow uncorrelated priors, which are either uniform or Gaussian distributed~\cite{Cui:2023jiw}.
    \item We multiply each form factor $\mathcal{F}^{i}_{\mathrm{LCSR}}(q^2)$ by an enhancement factor $W_{K^*} = 1.09 (1)$ to account for the finite $K^*$ width effect in $B\to K^*$ transition, as discussed in Ref.~\cite{Descotes-Genon:2019bud}.
     In the lattice QCD simulation, $K^*$ is a stable particle \cite{Horgan:2013hoa, Horgan:2015vla}. 
    \item In the high $q^2$ region, we reproduce the central values and correlation matrix of the lattice QCD results of the $B\to K^*$ form factors at three different points $q^2=\{12,14,16\}\,{\rm{GeV}}^2$ as well as physical-mass bottom quark and $2+1$ flavors of sea quarks. 
    To ensure the positive definiteness of the correlation matrix from the lattice QCD results, we modify the original matrix by adding an additional diagonal matrix of order $\mathcal{O}(10^{-6})$, namely $C_{\mathrm{latt}}=C_{\mathrm{latt, original}}+10^{-6}\mathrm{I}$.
    \item Taking into account the kinematic constraints,
    \begin{equation}
        \frac{m_B+m_V}{2m_V}A_1(0)-\frac{m_B-m_V}{2m_V}A_2(0)=A_0(0),\quad T_1(0)=T_2(0),
    \end{equation}
    we can derive the following exact relations between the expansion coefficients,
    \begin{equation}
        \frac{m^2_V}{m_B^2+m_V^2}b_0^{\mathcal{A}_1}-\frac{2m_B^2+m^2_V}{m_B^2+m_V^2}b_0^{\mathcal{A}_{12}}=b_0^{\mathcal{A}_0},\quad \frac{m_B^2}{m_B^2+m_V^2}b_0^{\mathcal{T}_{1}}=b_0^{\mathcal{T}_2}.
    \end{equation}
    \item We then construct
    \begin{equation}\label{chi2}
 \begin{aligned}
 \chi^2=& \sum_{i j}\left[\mathcal{F}_{\mathrm{LCSR}}^i\left(q^2\right)-\mathcal{F}_{\mathrm{fit}}^i\left(q^2 ; b_k^i\right)\right]\left(C_{\mathrm{LCSR}}^{-1}\right)_{i j}\left[\mathcal{F}_{\mathrm{LCSR}}^j\left(q^2\right)-\mathcal{F}_{\mathrm{fit}}^j\left(q^2 ; b_k^j\right)\right] \\
& \quad+\sum_{i j }\left[\mathcal{F}_{\mathrm{latt}}^i\left(q^2\right)-\mathcal{F}_{\mathrm{fit}}^i\left(q^2 ; b_k^i\right)\right]\left(C_{\mathrm{latt}}^{-1  }\right)_{i j}\left[\mathcal{F}_{\mathrm{latt}}^j\left(q^2\right)-\mathcal{F}_{\mathrm{fit}}^j\left(q^2 ; b_k^j\right)\right],
\end{aligned} 
\end{equation}
where $\mathcal{F}^i$ denote the central values of the form factors and $C_{ij}$ is the corresponding covariance matrix. We then extract the central values and the covariance of the coefficients $b_k^i$ by minimizing the $\chi^2$ function, yielding $\chi^2_{\text{min}}/\text{d.o.f}=40.1/23$. Our inputs as well as the fit results for the $z$-series coefficients, including the central values, uncertainties and all correlations, will be presented as supplemental material on the arXiv page.
\end{itemize}

\begin{figure}[!ht]
\begin{center}
\includegraphics[scale=0.5]{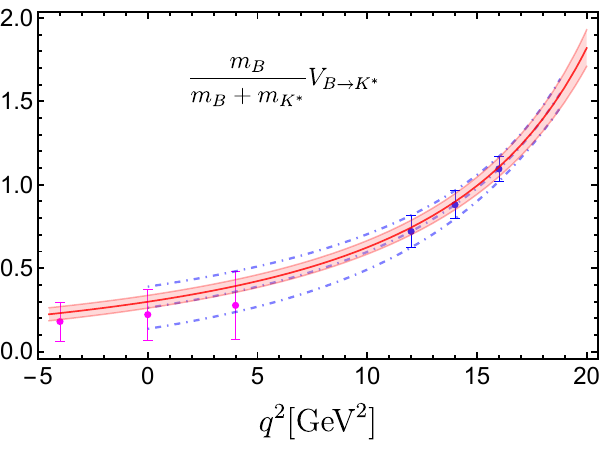}
\includegraphics[scale=0.5]{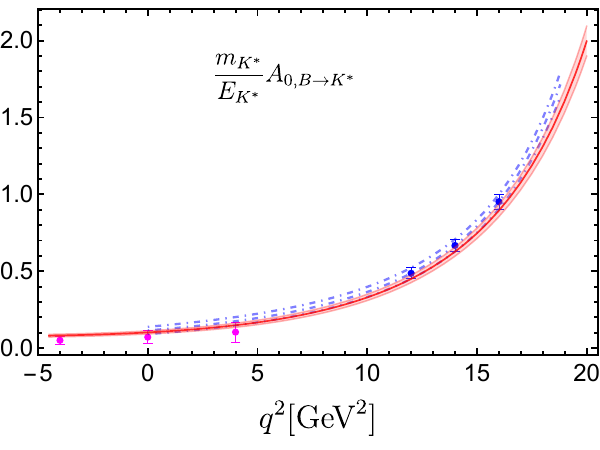}
\includegraphics[scale=0.5]{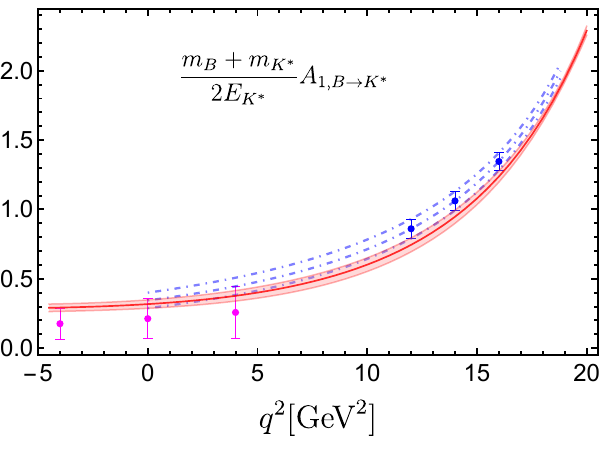}
\includegraphics[scale=0.5]{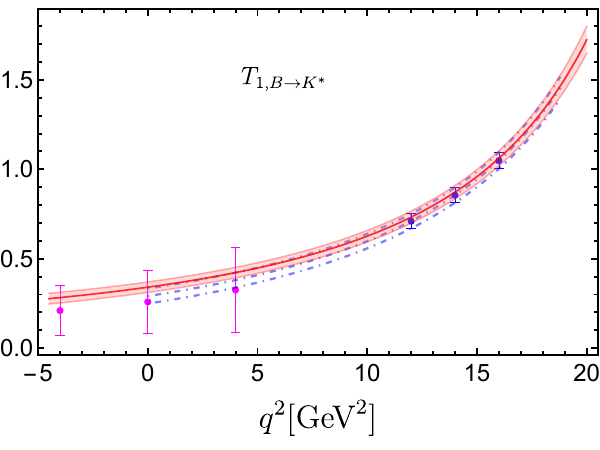}
\includegraphics[scale=0.5]{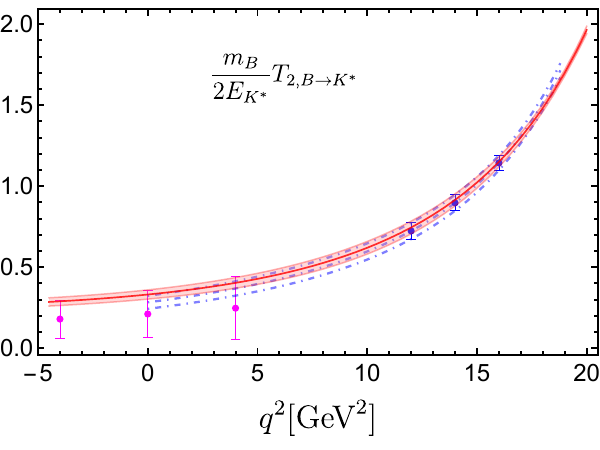}
\includegraphics[scale=0.5]{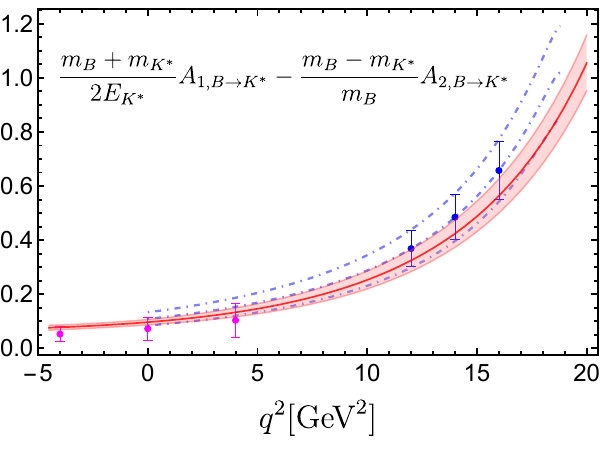}
\includegraphics[scale=0.5]{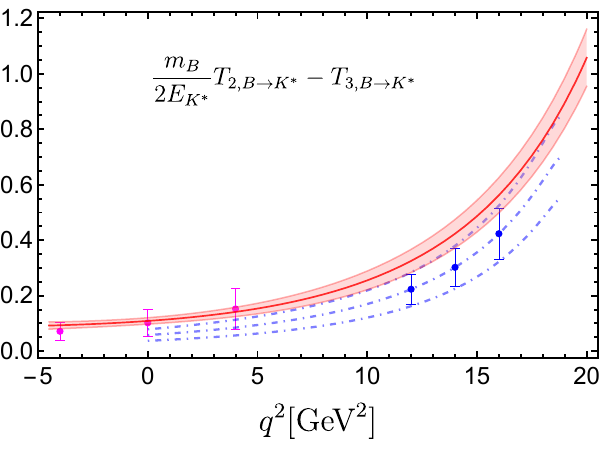}
\end{center}
\caption{Theoretical predictions of the $B\to K^*$ decay form factors (red band) obtained from the combined fit of updated LCSR (pink points) and lattice QCD (blue points) in the entire kinematical region. The `` lattice QCD only " predictions for these form factors are indicated by the blue dot-dashed line for a comparison.}

\label{fig:FFsq2}
\end{figure}

To further clarify the momentum-transfer dependence of the updated LCSR predictions and lattice QCD results, we present the combined fit results for the seven $B\to K^*$ form factors across the entire kinematic region in Fig.~\ref{fig:FFsq2}. BCL parametrization incorporates both our updated LCSR data (pink points) and lattice QCD data (blue points). For reference, lattice QCD predictions from prior studies are shown as a blue dot-dashed line. The inclusion of newly derived LCSR data in the low $q^2$ region substantially enhances the accuracy of theoretical predictions for the $B\to K^*$ form factors throughout the kinematic region, as demonstrated by the combined fit with lattice QCD simulations.

\subsection{Phenomenological analysis of the \texorpdfstring{$B\to K^* \,\nu_{\ell}\,\bar{\nu}_{\ell}$}{B to nunubar} observables}
The $B\to K^* \,\nu_{\ell}\,\bar{\nu}_{\ell}$ decays induced by the $b\to s$ flavor-changing neutral current (FCNC) represent one of the theoretically cleanest decay channels in heavy flavor physics. 
We now begin to explore the phenomenological implications of the newly determined $B\to K^*$ form factors for the electroweak penguin $B\to K^* \,\nu_{\ell}\,\bar{\nu}_{\ell}$ decays. 
Thanks to the high luminosity of the $\rm{Belle~II}$ experiment, the exclusive rare $B\to K^* \,\nu_{\ell}\,\bar{\nu}_{\ell}$ decays are expected to be observed with \(10 \, \text{ab}^{-1}\) of the data \cite{Belle-II:2018jsg, Halder:2021sgd} and the previous experimental measurements by $\rm{BaBar}$ \cite{BaBar:2013npw} and $\rm{Belle}$ \cite{Belle:2013tnz, Belle:2017oht} are also presented here. 
Notably, the  precision of the total branching fraction measurement for $B\to K^* \,\nu_{\ell}\,\bar{\nu}_{\ell}$  with \(50 \, \text{ab}^{-1}\) integrated luminosity is expected to reach approximately 10$\%$, rendering the experimental sensitivity comparable to the current theoretical uncertainty in Standard Model predictions.
Additionally, the longitudinal $K^*$ polarization fraction, which is highly sensitive to right-handed currents \cite{Belle-II:2018jsg}, is projected to be measured with an absolute uncertainty of 0.1, providing critical insights into potential beyond-Standard Model contributions.

We are therefore well motivated to further investigate the phenomenological aspects of the $B\to K^* \,\nu_{\ell}\,\bar{\nu}_{\ell}$ process, both to gain a deeper understanding of the strong interaction dynamics of the $B\to K^*$ form factors and to explore the potential role of exotic particles \(X\) in the context of dark matter, utilizing the form factors derived in this work. It is straightforward to derive the differential decay width formula \cite{Buras:2014fpa, Altmannshofer:2009ma}
\begin{align}
    \frac{d\Gamma (B\to K^{*} \,\nu_{\ell}\,\bar{\nu}_{\ell})}{dq^2}&=\frac{G^2_F \,\alpha^2_{em}}{256\,\pi^5}\frac{\lambda^{3/2}(m^2_B,m^2_{K^*}, q^2 )}{m^3_B\sin^4\theta_W}|\lambda_t|^2
    \bigg[X_t\bigg(\frac{m^2_t}{m^2_W}, \frac{m^2_H}{m^2_t},\sin\theta_W, \mu \bigg)\bigg]^2
    \nonumber\\
   &\times \bigg[H_V(q^2)+H_{A_1}(q^2)+H_{A_{12}}(q^2)\bigg],
\end{align}
where $\lambda(x,y,z)=x^2 + y^2 + z^2 - 2xy - 2xz - 2yz$ is the K\'allen function. The $\rm{CKM}$ matrix elements $\lambda_t=|V_{tb}V^{*}_{ts}|$ can be determined by the $\mathbf{UT}\,fit$ collaboration \cite{UTfit:2022hsi} and the input parameters appearing in the differential decay width are collected in Tab.~\ref{tab:para}. The short-distance Wilson coefficient $X_t$ can be expanded perturbatively in terms of the Standard Model coupling constants
\begin{align}
 X_t=X_t^{(0)}+\frac{\alpha_s}{4\pi}X_t^{\rm{QCD}(1)}+\frac{\alpha_{em}}{4\pi}X_t^{\rm{EW}(1)}+\cdots,
\end{align}
where the leading-order ($\rm{LO}$) contribution $X_t^{(0)}$ \cite{Inami:1980fz}, the next-to-leading-order  $\rm{(NLO)\, QCD}$  correction $X_t^{\rm{QCD}(1)}$ \cite{Buchalla:1992zm, Buchalla:1998ba, Misiak:1999yg} and the two-loop electroweak correction $X_t^{\rm{EW}(1)}$ \cite{Brod:2010hi} are already known analytically. We adopt $X_t=1.469$ in our work. The three invariant functions $H_i$ can be further expressed by the $B\to K^*$ form factors as
\begin{align}
 H_V(q^2)=&\frac{2q^2}{(m_B+m_{K^*})^2}[V(q^2)]^2,
 \nonumber\\
  H_{A_1}(q^2)=&\frac{2q^2(m_B+m_{K^*})^2}{\lambda(m^2_B,m^2_{K^*}, q^2 )}[A_1(q^2)]^2,
  \nonumber\\
  H_{A_{12}}(q^2)=&\frac{64m^2_Bm_{K^*}^2}{\lambda(m^2_B,m^2_{K^*}, q^2 )}[A_{12}(q^2)]^2,
\end{align}
with the helicity form factors $A_{12}$ \cite{Horgan:2013hoa}
\begin{align}
 A_{12}(q^2)=\frac{(m_B+m_{K^*})^2(m^2_B-m_{K^*}^2-q^2)A_1(q^2)-\lambda(m^2_B,m^2_{K^*}, q^2 )A_2(q^2)}{16m_Bm_{K^*}^2(m_B+m_{K^*})}.
\end{align}
\begin{figure}[!htb]
\begin{center}
\includegraphics[width=0.75\columnwidth]{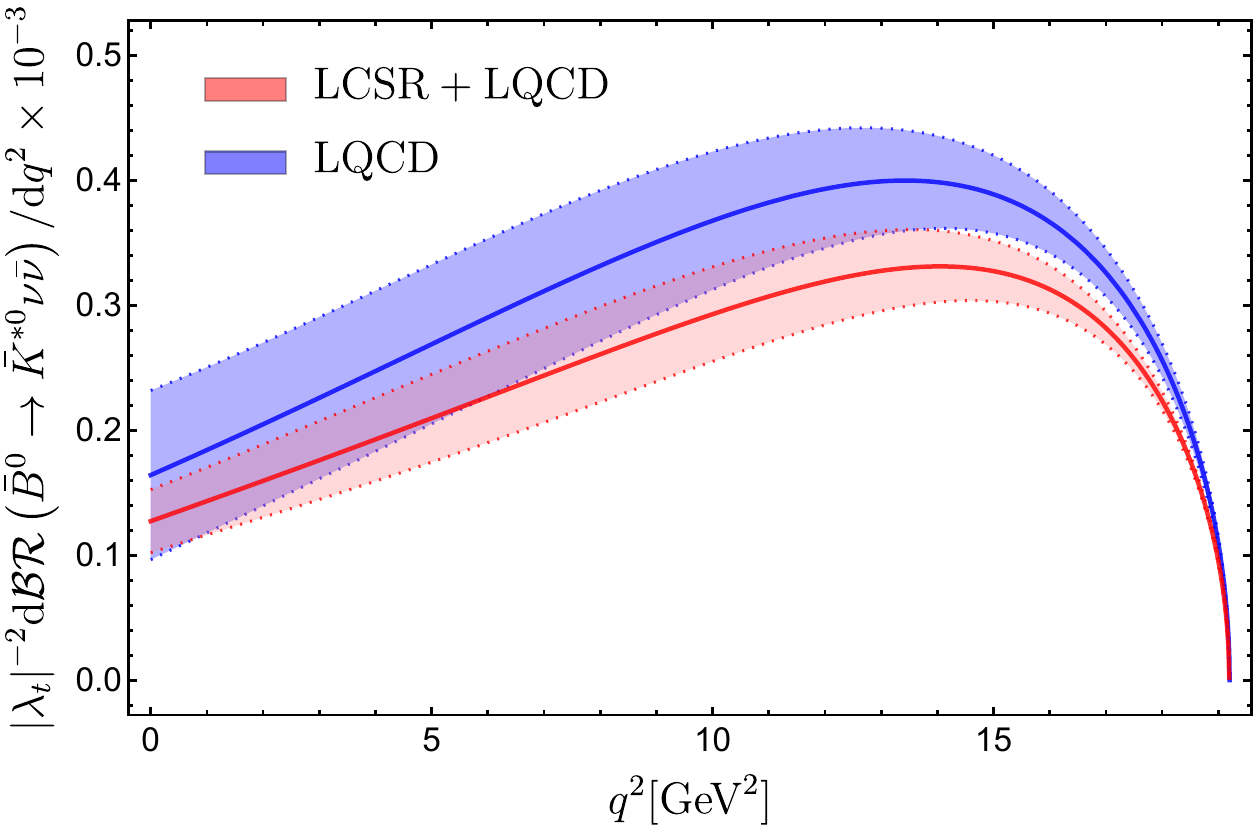}
\end{center}
\vspace{-0.5cm}
\caption{ Theory predictions for the CKM-independent differential branching fraction of $B^0\to K^{*0} \,\nu_{\ell}\,\bar{\nu}_{\ell}$  
by applying the form factors determined from the combined fits (pink band) and from the lattice simulations (blue band) \cite{Horgan:2013hoa, Horgan:2015vla}.}
\label{fig:FFsBr}
\end{figure}
To probe new physics effects beyond the SM, $\lambda_t$ is typically determined through CKM unitarity. However, inconsistencies persist in the extracted values of the CKM matrix element $V_{cb}$ across different processes.
For future studies, we present the CKM-independent branching ratio $|\lambda_t|^{-2}\mathcal{BR} (B^0\to K^{*0}\nu_{\ell}\,\bar{\nu}_{\ell})$ estimated with various strategies in Tab.~\ref{BR}. The branching ratios derived from updated $B\to K^*$ form factors agree within 2.5$\sigma$ with the results obtained from sum rules based on $K^*$ distribution amplitudes~\cite{Bharucha:2015bzk}.
Additionally, in Fig.~\ref{fig:FFsBr},  we display our theoretical prediction for the CKM-independent differential branching fraction of $B^0\to K^{*0} \,\nu_{\ell}\,\bar{\nu}_{\ell}$ and show the result from lattice QCD calculations for comparison. It is evident that the combined fit result exhibits significantly smaller uncertainty than lattice QCD predictions across the entire momentum region.
 Finally, our numerical results for the differential branching ratio of $B\to K^{*} \,\nu_{\ell}\,\bar{\nu}_{\ell}$ are summarized in Tab.~\ref{tab:BR},  where we have adopt $\lambda_t=41.25\times 10^{-3}$. The total uncertainty is dominated by the uncertainties in the hadronic form factors. 

\begin{table}[htb]
\centering \setlength\tabcolsep{8.0pt} \def\arraystretch{1.5}
\caption{The CKM-independent branching ratio of $B^0\to K^{*0} \,\nu_{\ell}\,\bar{\nu}_{\ell}$ process from updated form factors (left), lattice QCD FFs (middle) and LCSRs with $K^*$-meson LCDAs (right). }
\vspace{5pt}
\begin{tabular}{|l|c|c|c|}
\hline
$10^{-3}\times (\lambda_t)^{-2}\mathcal{BR}$ & This work & Ref.~\cite{Horgan:2013hoa, Horgan:2015vla}& Ref.~\cite{Bharucha:2015bzk}  \\ \hline
 $ B^0\to K^{*0}\nu_{\ell}\,\bar{\nu}_{\ell} $& 4.76(56)  & 5.86(93) & 5.85(58) \\ \hline
\end{tabular}
\label{BR}
\end{table}
\begin{table}[htb] 
\centering \setlength\tabcolsep{2.6pt} \def\arraystretch{1.5}
\caption{Theory predictions for the integrated differential observables $\Delta\mathcal{BR}(q^2_1, q^2_2)$, and $\Delta F_L (q^2_1, q^2_2)$ obtained from the exclusive $B\to K^*$ form factors and additional long-distance contribution.}
\vspace{5pt}
\begin{tabular}{| c| c |c |c |} 
\hline 
$[q^2_1, q^2_2] \,(\rm{in}\, \rm{GeV}^2)$ & $10^6\times \Delta\mathcal{BR}^{B^0\to K^{*0} \,\nu_{\ell}\,\bar{\nu}_{\ell}}(q^2_1, q^2_2)$ & $10^6\times\Delta\mathcal{BR}^{B^+\to K^{*+} \,\nu_{\ell}\,\bar{\nu}_{\ell}}(q^2_1, q^2_2)$ & $\Delta F_L (q^2_1, q^2_2)$ \\
\hline 
$[0.0, 1.0]$ & 0.23(5) & 0.33(5)
 &0.93(2)\\
$[1.0, 2.5]$ & 0.40(7) & 0.55(8)
 &0.79(4)\\
$[2.5, 4.0]$ & 0.46(8)  & 0.61(9)
 &0.67(5)\\
$[4.0, 6.0]$ & 0.71(12) & 0.91(13)
 &0.57(5)\\
$[6.0, 8.0]$ & 0.83(13) & 1.03(14)
 &0.48(5)\\
$[8.0, 12.0]$ & 1.99(26) & 2.39(28)
 &0.40(4)\\
$[12.0, 16.0]$ & 2.22(20) & 2.61(22)
 &0.33(2)\\
$[16.0, (m_B-m_{K^*})^2]$ & 1.26(6) & 1.53(7) 
 &0.31(1)\\
$[0.0, (m_B-m_{K^*})^2]$ & 8.09(96) & 9.95(1.05)
 &0.44(4)\\
\hline 
\end{tabular} 
\label{tab:BR}
\end{table}
There is an additional long-distance (LD) contribution to the counterpart channel $B^+\to K^{*+} \,\nu_{\ell}\,\bar{\nu}_{\ell}$ involving a charged $B$-meson due to the double-charged current interaction $B^+\to\tau^+(\to K^{*+}\nu_{\tau})\bar{\nu}_{\tau}$, as originally discussed in Ref.~\cite{Kamenik:2009kc}. In the narrow $\tau$-lepton width limit, we express the tree-level LD contribution to the differential decay rate  
\begin{equation}
\begin{aligned}
  \frac{d\Gamma (B^+\to K^{*+} \,\nu_{\ell}\,\bar{\nu}_{\ell})}{dq^2}\Big|_{\text{LD}}&=
  \frac{G^4_F|V_{ub}V_{us}^*|^2}{64\pi^2m^3_{B}}|f_{B}f_{K^{*}}|^2\frac{m_\tau^3}{\Gamma_\tau} \\
  &\times 
 [(m^2_{B}-m_{\tau}^2)(m_{\tau}^2-m^2_{K^{*}})-(m_\tau^2-2m^2_{K^{*}})q^2].
\end{aligned}
\end{equation}
This long-distance effect arising from weak annihilation mediated by the on-shell $\tau$-lepton accounts for approximately 10\% of the electroweak penguin amplitude, which is numerically significant for the charged channel $B^+\to K^{*+} \,\nu_{\ell}\,\bar{\nu}_{\ell}$.
Moreover, the interference effect between the tree and penguin
amplitudes turns out to be negligible numerically due to extremely small width of $\tau$-lepton~\cite{Kamenik:2009kc}.

We then proceed to define the  differential longitudinal $K^*$
polarization fraction $F_L$ of the electroweak penguin decays
$B\to K^* \,\nu_{\ell}\,\bar{\nu}_{\ell}$
\begin{align}
 F_L(q^2)=\frac{H_{A_{12}}(q^2)}{H_V(q^2)+H_{A_1}(q^2)+H_{A_{12}}(q^2)}.
\end{align}
In addition, we introduce two $q^2$-binned observables for comparison with future high-luminosity $\rm{Belle~II}$ data
\begin{equation}
\begin{aligned}
 &\Delta\mathcal{BR}(q^2_1, q^2_2)=\tau_{B}\int^{q^2_2}_{q^2_1}dq^2 \frac{d\Gamma (B\to K^{*} \,\nu_{\ell}\,\bar{\nu}_{\ell})}{dq^2},
 \\
 &\Delta F_L(q^2_1, q^2_2)=\frac{\int^{q^2_2}_{q^2_1}dq^2 \lambda^{3/2}(m^2_B,m^2_{K^*}, q^2 ) H_{A_{12}}(q^2)}{\int^{q^2_2}_{q^2_1}dq^2 \lambda^{3/2}(m^2_B,m^2_{K^*}, q^2 )[H_V(q^2)+H_{A_1}(q^2)+H_{A_{12}}(q^2)]}.
\end{aligned}
\end{equation}
Our predictions for these observables,  with the choice of $q^2$-intervals following \cite{Belle-II:2018jsg}, are summarized in Tab.~\ref{tab:BR}.
The theoretical uncertainties of the $q^2$-binned longitudinal $K^*$ polarization fractions, $\Delta F_L$,  are significantly smaller than those of the branching ratio predictions,  $\Delta\mathcal{BR}$, due to the reduced sensitivity of the form-factor ratios to the precise shapes of the $B$-meson distribution amplitudes.

\section{Summary}

In this work, we have comprehensively investigated subleading-power corrections to the $B\to K^*$ form factors {up to twist-six} within the framework of light-cone sum rules (LCSR) with $B$-meson LCDAs. 
The corrections arise from two-particle and three-particle $B$-meson higher-twist light-cone distribution amplitudes,  power-suppressed terms in the expansion of the strange quark propagator, the subleading-power effective current $\bar{q}\Gamma[i\slashed{D}_{\perp}/(2m_b)]h_v$ in HQET, and the four-particle twist-five and twist-six $B$-meson LCDAs in the factorization approximation.
Incorporating the leading-power contribution at NLL accuracy from Ref.~\cite{Gao:2019lta} with our newly derived NLP contributions, we ultimately obtain updated predictions for the $B\to K^*$ form factors with SCET sum rules in the large recoil region. It is shown that power corrections account for approximately a 30\% correction to the tree-level result, which is comparable to the NLL contribution.
Moreover, we reach a similar conclusion that the dominant source of power corrections arises from the two-particle higher-twist $B$-meson LCDAs as in Ref.~\cite{Gubernari:2018wyi}, while the impact of the four-particle corrections is numerically insignificant.

We employ a three-parameter model to describe the $B$-meson LCDAs and adopt the conventional inverse moment $\lambda_B$ = 350(150)~MeV. 
{In addtion, we estimate the $B\to K^*$ form factors  by adopting the inverse moment $\lambda_B =389(35)$~MeV from recent lattice QCD  calculations for comparison.}
The dominant uncertainties in the form factors originate from the inverse moments $\lambda_B$ and $\{\sigma_1, \sigma_2 \}$. 
{In the future, lattice QCD studies may reduce the uncertainties of the inverse moments by systematically investigating subleading-power contributions via this first principle approach.}
By adopting the BCL parameterization and performing a combined fit of the newly derived LCSR predictions in the low $q^2$ region and the lattice QCD results in the high $q^2$ region, we extrapolate the $B\to K^*$ form factors to the entire momentum region. We find that the $B\to K^*$ form factors derived from the combined fits exhibit smaller uncertainties than those obtained solely using lattice data points. 
Furthermore, the combined fits to the LCSR and lattice QCD inputs for the $B\to K^*$ form factors not only provide predictions for the form factors that are applicable across the entire kinematic range, but also confirm the consistency of the two complementary methods by ensuring their agreement at intermediate $q^2$ values. 
Having at our disposal the $B\to K^*$ form factors in the entire momentum region, we proceed to predict the differential decay widths for $B\to K^* \,\nu_{\ell}\,\bar{\nu}_{\ell}$ processes, including long-distance effects in the charged $B\to K^* \,\nu_{\ell}\,\bar{\nu}_{\ell}$ decay. 
We present the CKM-independent differential branching ratio of $B^0\to K^{*0} \,\nu_{\ell}\,\bar{\nu}_{\ell}$ obtained from the combined fits, alongside the lattice simulation result for comparison. The results with different fit data inputs are consistent with each other and the combined fits to the differential branching ratio of $B^0\to K^{*0} \,\nu_{\ell}\,\bar{\nu}_{\ell}$ yield smaller uncertainties. Finally, we obtain the branching ratios  $\mathcal{BR}(\bar{B}^0 \to \bar{K}^{*0} \nu_\ell\bar{\nu}_\ell)=8.09(96)\times 10^{-6}$, $\mathcal{BR}(\bar{B}^+ \to \bar{K}^{*+} \nu_\ell\bar{\nu}_\ell)=9.95(1.05)\times 10^{-6}$, and the longitudinal $K^*$ polarization fraction $F_L=0.44(4)$.

For the $b\to s$ induced flavor-changing neutral current processes, a crucial task is to improve the precision of the $B\to K^*$ form factors. 
It can be further improved with respect to the following three aspects: developing model-independent methods for accurately describing the $B$-meson light-cone distribution amplitudes, reducing the uncertainties in  the non-perturbative input parameters, such as the inverse moment $\lambda_B$ of the leading-twist $B$-meson LCDA, 
extending calculations to next-to-next-to-leading order (NNLO) at leading power and next-to-leading order (NLO) at subleading power, and
imposing stricter unitarity bounds on the 
$z$-series parameterizations to further constrain uncertainties. 

Ultimately, we emphasize that our improved $B\to K^*$ form factors are crucial for investigating flavor-changing neutral current processes and determining the branching ratios of the electroweak penguin processes $B\to K^* \,\nu_{\ell}\,\bar{\nu}_{\ell}$. With the high luminosity $\rm{Belle~II}$ experimental data upcoming, our predictions will be further tested in the future.

\section*{Acknowledgement}

We thank Shan Cheng for valuable discussions. Yue-Long Shen acknowledges support from the National Natural Science Foundation
of China with Grant No. 12175218, and the Natural Science
Foundation of Shandong Province under the Grant No. ZR2024MA076. Dong-Hao Li acknowledges support from the National Natural Science
Foundation of China with Grant No. 12447154.
The work of UGM was supported in part  by the CAS President's International Fellowship Initiative (PIFI) under Grant No.~2025PD0022.

\appendix
\section{Hard function for the SCET curents at \texorpdfstring{$\mathcal{O}(\alpha_s)$}{} }\label{hard function}
Here we present the hard coefficient functions of A0-type and B1-type~$\rm{SCET}_I$~currents in~$B\to K^*$~form factors up to~$\mathcal{O}(\alpha_s)$
\begin{align}
C^{(\rm{A0})}_{f_+}&=1+\frac{\alpha_sC_F}{4\pi}\bigg\{-2\ln^2(\frac{r}{\hat{\mu}})+5\ln(\frac{r}{\hat{\mu}})-2\text{Li}_2(1-r)-3\ln r-\frac{\pi^2}{12}-6\bigg\},\\
 C^{(\rm{A0})}_{f_0}&=1+\frac{\alpha_sC_F}{4\pi}\bigg\{-2\ln^2(\frac{r}{\hat{\mu}})+5\ln(\frac{r}{\hat{\mu}})-2\text{Li}_2(1-r)-\frac{3-5r}{1-r}\ln r-\frac{\pi^2}{12}-4\bigg\},\\
 C^{(\rm{A0})}_{f_T}&=1+\frac{\alpha_sC_F}{4\pi}\bigg\{-2\ln\hat{\nu}-2\ln^2(\frac{r}{\hat{\mu}})+5\ln(\frac{r}{\hat{\mu}})-2\text{Li}_2(1-r)-\frac{3-r}{1-r}\ln r-\frac{\pi^2}{12}-6\bigg\},\\
 C^{(\rm{A0})}_{V}&=1+\frac{\alpha_sC_F}{4\pi}\bigg\{-2\ln^2(\frac{r}{\hat{\mu}})+5\ln(\frac{r}{\hat{\mu}})-2\text{Li}_2(1-r)-\frac{3-2r}{1-r}\ln r-\frac{\pi^2}{12}-6\bigg\},\\
  C^{(\rm{A0})}_{T_1}&=1+\frac{\alpha_sC_F}{4\pi}\bigg\{-2\ln\hat{\nu}-2\ln^2(\frac{r}{\hat{\mu}})+5\ln(\frac{r}{\hat{\mu}})-2\text{Li}_2(1-r)-3\ln r-\frac{\pi^2}{12}-6\bigg\},\\
  C^{( \rm{B1})}_{f_+}&=(-2+1/r)+\mathcal{O}(\alpha_s),\qquad
  C^{( \rm{B1})}_{f_0}=(-1/r)+\mathcal{O}(\alpha_s),\\
  C^{( \rm{B1})}_{f_T}&=(1/r)+\mathcal{O}(\alpha_s),\qquad
  C^{( \rm{B1})}_{V}=0+\mathcal{O}(\alpha_s),\qquad
  C^{(\rm{B1})}_{T_1}=-1+\mathcal{O}(\alpha_s), 
\end{align}
where we introduced three variables
\begin{align}
 r=\frac{n\cdot p}{m_b},\qquad 
 \hat{\mu}=\frac{\mu}{m_b},\qquad
 \hat{\nu}=\frac{\nu}{m_b}.
\end{align}
\section{Effective \texorpdfstring{$B$}{}-meson distribution amplitudes}\label{effective-BDA}
For brevity, we introduce the effective $B$-meson distribution amplitudes $\phi^-_{B,\text{eff}}$, $\tilde{\phi}^-_{B,\text{eff}}$ and $\phi^+_{B,m}$ absorbing the hard-collinear fluctuations,
\begin{align}
 &\phi_{B,\text{eff}}^{-}(\omega',\mu)= \phi_{B}^{-}(\omega',\mu)+\frac{\alpha_s C_F}{4\pi}\bigg\{\int_{0}^{\omega'}d\omega\bigg[\frac{2}{\omega-\omega'}\bigg(\ln\frac{\mu^2}{n\cdot p\omega'}
 -2\ln\frac{\omega'-\omega}{\omega'}\bigg)\bigg]_{\oplus}\phi_{B}^{-}(\omega,\mu)
 \nonumber\\
 -&\int_{\omega'}^{\infty}d\omega\bigg[\ln^2
 \frac{\mu^2}{n\cdot p\omega'}-\bigg(2\ln\frac{\mu^2}{n\cdot p\omega'}+3\bigg)\ln\frac{\omega-\omega'}{\omega'}
 +2\ln\frac{\omega}{\omega'}+\frac{\pi^2}{6}-1\bigg]\frac{d\phi_{B}^{-}(\omega,\mu)}{d\omega}
 \bigg\},
 \\
  &\tilde{\phi}_{B,\text{eff}}^{-}(\omega',\mu,\nu)=\phi_{B}^{-}(\omega',\mu)+\frac{\alpha_s C_F}{4\pi}\bigg\{\int_{0}^{\omega'}d\omega\bigg[\frac{2}{\omega-\omega'}\bigg(\ln\frac{\mu^2}{n\cdot p\omega'}-2\ln\frac{\omega'-\omega}{\omega'}-\frac{1}{2}\bigg)\bigg]_{\oplus}\phi_{B}^{-}(\omega,\mu)
\nonumber \\ 
-&\int_{\omega'}^{\infty}d\omega\bigg[\ln^2
 \frac{\mu^2}{n\cdot p\omega'}-\ln
 \frac{\nu^2}{n\cdot p\omega'}-\bigg(2\ln\frac{\mu^2}{n\cdot p\omega'}+4\bigg)\ln\frac{\omega-\omega'}{\omega'}+2\ln\frac{\omega}{\omega'}+\frac{\pi^2}{6}-1\bigg]\frac{d\phi_{B}^{-}(\omega,\mu)}{d\omega}
 \bigg\},
  \\
 &\phi_{B,m}^{+}(\omega',\mu)=\frac{\alpha_s C_F}{4\pi}m_q\int_{\omega'}^{\infty}d\omega\ln\frac{\omega-\omega'}{\omega'}\frac{d}{d\omega}\frac{\phi_{B}^{+}(\omega,\mu)}{\omega},
\end{align}
the plus function appeared in the above equations is defined by
\begin{align}
\int_{0}^{\omega'}d\omega[f(\omega,\omega')]_{\oplus}g(\omega)
 =\int_{0}^{\omega'}d\omega f(\omega,\omega')[g(\omega)-g(\omega')].
\end{align}
\section{Dispersion integral formulas}\label{dis func-app}
After Borel transformation, the dispersion functions appearing in the factorization formulas of $B\to K^*$ form factors are listed in the following
\begin{equation}
    \begin{aligned}
{f}_{2,1}[\phi(\omega)]=&-\int^{\omega_s}_{0}d\omega 
        e^{-\frac{\omega}{\omega_M}}\phi(\omega),\\
{f}_{2,2}[\phi(\omega)]=&e^{-\frac{\omega_s}{\omega_M}}\phi(\omega_s)+
      \int^{\omega_s}_{0}d\omega\frac{e^{-\frac{\omega}{\omega_M}}}{\omega_M}\phi(\omega),\\
 {f}_{3,2}[\phi(\omega_1,\omega_2,u)]=&e^{-\frac{\omega_s}{\omega_M}}
      \int^{\omega_s}_{0}d\omega_1\int^{\infty}_{\omega_s-\omega_1}\frac{d\omega_2}{\omega_2}\phi(\omega_1,\omega_2,\frac{\omega_s-\omega_1}{\omega_2})\\
   &+\int^{\omega_s}_{0}d\omega\int^{\omega}_{0}d\omega_1
     \int^{\infty}_{\omega-\omega_1}\frac{d\omega_2}{\omega_2}
     \frac{e^{-\frac{\omega}{\omega_M}}}{\omega_M}
      \phi(\omega_1,\omega_2,\frac{\omega-\omega_1}{\omega_2}),\\
{f}_{3,3}[\phi(\omega_1,\omega_2,u)]=&
      -\frac{1}{2}e^\frac{-\omega_s}{\omega_M}
      \left\{
      \int^{\infty}_{\omega_s}\frac{d\omega_2}{\omega_2}
      \phi(0,\omega_2,\frac{\omega_s}{\omega_2})
\right.\\
      &\left. +\int^{\omega_s}_{0}
      d\omega_1
     \int^{\infty}_{\omega_s-\omega_1}\frac{d\omega_2}{\omega_2}\left(\frac{d}{d\omega_1}+\frac{1}{\omega_M}\right)\phi(\omega_1,\omega_2,\frac{\omega_s-\omega_1}{\omega_2})\right\}
\\
    &-\frac{1}{2\omega_M^2}\int^{\omega_s}_{0}d\omega
    \int^{\omega}_{0}
    d\omega_1
   \int^{\infty}_{\omega-\omega_1}\frac{d\omega_2}{\omega_2}e^{\frac{-\omega}{\omega_M}}
     \phi(\omega_1,\omega_2,\frac{\omega-\omega_1}{\omega_2}),
    \end{aligned}
\end{equation}
where $\phi$ stands for the general $B$-meson LCDAs or their combinations appearing in the function $f_{i,j}$. The function $f_{i,j}$ describes the contribution of terms in the form $\phi(\omega)/(\omega-\cdots)^j$, with $\phi(\omega)$ being the $i$-particle LCDA and the denominator raised to the $j$-th power. 

\section{Modeling the  \texorpdfstring{$B$}{}-meson LCDAs}\label{B meson LCDAs}
The general ansatz for leading- and higher-twist $B$-meson LCDAs at the reference scale $\mu_0=1$ GeV~\cite{Beneke:2018wjp} can be systematically established
in such a way that both tree-level
equations of motion constraints and the normalization conditions of the 
LCDAs~\cite{Grozin:1996pq, Braun:2017liq} are satisfied. 
\begin{align}
\phi_B^+(\omega) &= \omega \, \mathbb{F}(\omega;-1) \,, \qquad\quad
\phi_B^{-\rm WW}(\omega) = \mathbb{F}(\omega;0) \,,
\nonumber \\
\phi_B^{-\rm t3}(\omega) &= \frac{1}{6}\,\mathcal{N} \, 
(\lambda^2_E - \lambda^2_H ) \,
\Big[
-\omega^2\,\mathbb{F}(\omega;-2)
+ 4\,\omega\,\mathbb{F}(\omega;-1) -2\, \mathbb{F}(\omega;0)
\Big] \,,
\nonumber \\
\phi_3(\omega_1,\omega_2) &=  \frac{1}{2}\,\mathcal{N} \, 
(\lambda^2_E - \lambda^2_H ) \, 
\omega_1\,\omega^2_2\, \mathbb{F}(\omega_1+\omega_2;-2) \,,
\nonumber \\
\widehat{g}_B^+(\omega) &=
\frac{1}{4}\bigg[
2\,\omega\,(\omega-\bar{\Lambda})\,\mathbb{F}(\omega;0)
+(3\omega-2\bar{\Lambda})\,\mathbb{F}(\omega;1)
+3\,\mathbb{F}(\omega;2)
\nonumber \\
& -\frac{1}{6} \, \mathcal{N} \, (\lambda^2_E - \lambda^2_H ) \, 
\omega^2 \,\mathbb{F}(\omega;0)\bigg]\,,
\nonumber \\
\widehat{g}_B^-(\omega) &=
\frac{1}{4}\,\bigg\{ (3\omega-2\bar{\Lambda})\,\mathbb{F}(\omega;1)
+3\, \mathbb{F}(\omega;2)
\nonumber \\
& +\frac{1}{3} \, \mathcal{N} \, 
(\lambda^2_E - \lambda^2_H ) \, 
\omega\,\bigg[
\omega\, (\bar{\Lambda}-\omega) \, \mathbb{F}(\omega;-1)
-\Big(2\,\bar{\Lambda} -\frac{3}{2}\,\omega \Big) \,
\mathbb{F}(\omega;0) \bigg]\bigg\}\,,
\nonumber \\
\phi_4(\omega_1,\omega_2) &= \frac{1}{2}\,\mathcal{N} \, 
(\lambda^2_E + \lambda^2_H ) \, 
\omega^2_2\, \mathbb{F}(\omega_1+\omega_2;-1) \,,
\nonumber \\
\psi_4(\omega_1,\omega_2) 
&= \mathcal{N} \, 
\lambda^2_E  \, 
\omega_1\, \omega_2\,  \mathbb{F}(\omega_1+\omega_2;-1) \,,\qquad
\widetilde{\psi}_4(\omega_1,\omega_2) = \mathcal{N} \, 
\lambda^2_H  \, 
\omega_1\, \omega_2\,  \mathbb{F}(\omega_1+\omega_2;-1) \,,
\nonumber \\
\phi_5(\omega_1,\omega_2) &= \mathcal{N} \, 
(\lambda^2_E + \lambda^2_H ) \, 
\omega_1\, \mathbb{F}(\omega_1+\omega_2;0) \, , \hspace{2.7mm}
\psi_5(\omega_1,\omega_2) = - \mathcal{N} \, 
\lambda^2_E \, 
\omega_2\,  \mathbb{F}(\omega_1+\omega_2;0) \,,
\nonumber \\
\widetilde{\psi}_5(\omega_1,\omega_2) &= - \mathcal{N} \, 
\lambda^2_H \, 
\omega_2\,  \mathbb{F}(\omega_1+\omega_2;0) \,,
\nonumber \\
\phi_6(\omega_1,\omega_2) &= \mathcal{N} \, 
(\lambda^2_E - \lambda^2_H ) \, \mathbb{F}(\omega_1+\omega_2;1)\,,\label{model:ansatz}
\end{align}
where 
\begin{align}
\mathcal{N} &= \frac{1}{3}\,\frac{\beta\,(\beta+1)}{\alpha\,(\alpha+1)}\,
\frac{1}{\omega^2_0} \, ,\qquad\qquad\qquad
\bar{\Lambda} = \frac{3}{2}\, \frac{\alpha}{\beta} \, \omega_0\,,
\notag\\
\mathbb{F}(\omega;n) &\equiv 
\omega^{n-1}_0 \,  
U(\beta-\alpha,2-n-\alpha,\omega/\omega_0) \,
\frac{\Gamma(\beta)}{\Gamma(\alpha)}\,
e^{-\omega/\omega_0} \,, \label{model:bb-F}
\end{align}
with $U(a, b, z)$ is the hypergeometric $U$ function.
The appearing HQET parameters $\lambda_E^2$ and $\lambda_H^2$ at the reference scale $\mu_0=1$ GeV are defined by the matrix element of local quark-gluon-quark operator,
\begin{align}
\langle 0| \bar q(0) g_s G_{\mu\nu}(0)\Gamma h_v(0)|\bar B(v)\rangle &=-\frac{i}{6} \widetilde{f}_B(\mu)m_B \lambda^2_H {\rm{Tr}}\Big[\gamma_5\Gamma P_+ \sigma_{\mu\nu}\Big]
\label{def:lambdaEH}\\
&-\vspace*{3mm}\frac{1}{6} \widetilde{f}_B(\mu)m_B\Big( \lambda^2_H- \lambda^2_E\Big)
  {\rm{Tr}}\Big[\gamma_5\Gamma P_+(v_\mu\gamma_\nu-v_\nu\gamma_\mu)\Big]\,.\notag
\end{align}
The matrix element can be estimated adopting QCD sum rules yielding, 
\begin{align}
\lambda^2_E = 0.11\pm 0.06~\text{GeV}^2, && 
\lambda^2_H =  0.18\pm 0.07~\text{GeV}^2,&& \text{\cite{Grozin:1996pq}}
\\
\lambda^2_E = 0.03\pm 0.02~\text{GeV}^2, && 
\lambda^2_H = 0.06\pm 0.03~\text{GeV}^2, &&  \text{\cite{Nishikawa:2011qk}}
\\
\lambda^2_E = 0.01\pm 0.01~\text{GeV}^2, && 
\lambda^2_H = 0.15\pm 0.05~\text{GeV}^2, &&  \text{\cite{Rahimi:2020zzo}}
\label{QCDSR:lambdaEH}
\end{align}
where we take into account that the method used to estimate $\lambda^2_E$ and $\lambda^2_H$, as discussed in Ref.~\cite{Rahimi:2020zzo}, unfortunately not only disrupts the convergence of the operator-product-expansion (OPE), but  also enhances the contributions from the continuum and higher excited states. Therefore, we will use the numerical results of $\lambda^2_E$ and $\lambda^2_H$ from the Tab.~\ref{tab:para}, which can cover the ranges allowed by Refs.~\cite{Grozin:1996pq, Nishikawa:2011qk}, and simultaneously satisfy the upper bounds imposed by Ref.~\cite{Rahimi:2020zzo}.

We expect that the model for $B$-meson light-cone distribution amplitudes is only valid in the small momenta region and the inverse logarithmic moments are only sensitive to the small momentum behavior of the distribution amplitude. Employing the definitions of inverse logarithmic moments of the leading-twist $B$-meson LCDA
\begin{align}
\frac{1}{\lambda_B(\mu)} &=  
\int^\infty_0 \frac{d\omega}{\omega}\,\phi_B^+(\omega,\mu) \,,
\nonumber \\
\frac{\widehat{\sigma}_n(\mu)}{\lambda_B(\mu)} &= 
\int^\infty_0 \frac{d\omega}{\omega}\,
\ln^n\frac{e^{-\gamma_E}\lambda_B(\mu)}{\omega}\,
\phi_B^+(\omega,\mu) \, ,
\label{eq:log-moms}
\end{align}
we can determine the parameters of  three-parameter model for  $B$-meson light-cone distribution amplitudes in Eq.~\eqref{model:ansatz}
\begin{align}
\lambda_B(\mu) &= \frac{\alpha-1}{\beta-1}\,\omega_0\,,
\nonumber \\
\widehat \sigma_1(\mu) &= 
\psi(\beta-1) -\psi(\alpha-1) + \ln\frac{\alpha-1}{\beta-1}\,,
\nonumber \\
\widehat \sigma_2(\mu)&= 
\widehat \sigma_1^2(\mu)+\psi'(\alpha-1)-\psi'(\beta-1)
+\frac{\pi^2}{6},
\label{eq:hatsigma}
\end{align}
where $\gamma_E$ and $\psi(x)$ denote the Euler-Mascheroni constant and the digamma function, respectively.
The scale dependence of these moments at one-loop level is given by
\begin{align}
\frac{\lambda_B(\mu_0)}{\lambda_B(\mu)} 
&= 1+ \frac{\alpha_s(\mu_0)\,C_F}{4\pi}\ln\frac{\mu}{\mu_0}
\Big[2-2\ln\frac{\mu}{\mu_0} -4\sigma_1(\mu_0) \Big] \,,
\nonumber \\
\widehat{\sigma}_1(\mu) 
&= \widehat{\sigma}_1(\mu_0)
+ \frac{\alpha_s(\mu_0)\,C_F}{4\pi}\,4\ln\frac{\mu}{\mu_0}
\Big[\widehat{\sigma}^2_1(\mu_0)-\widehat{\sigma}_2(\mu_0) \Big] \,.
\end{align}
 Then we construct the LL resummation (evolution) for the twist-2 and 3 two-particle $B$-meson LCDAs.
The explicit expressions can be found in Ref.~\cite{Beneke:2018wjp}
\begin{align}
\phi_B^+(\omega,\mu)
=&~  U_\phi(\mu,\mu_0) \,  \frac{1}{\omega^{p+1}}\,
\frac{\Gamma(\beta)}{\Gamma(\alpha)} \,
\mathcal{G}(\omega;0,2,1)\,,
\nonumber \\
\phi^{-\rm WW}_B(\omega,\mu)
=&~   U_\phi(\mu,\mu_0) \,  \frac{1}{\omega^{p+1}}\,
\frac{\Gamma(\beta)}{\Gamma(\alpha)} \,
\mathcal{G}(\omega;0,1,1)\,,
\nonumber \\
\phi^{-\rm t3}_B(\omega,\mu)
=&  - \frac{1}{6}\, U^{\rm t3}_\phi(\mu,\mu_0) \,\mathcal{N} \, 
(\lambda^2_E - \lambda^2_H ) \,\frac{\omega^2_0}{\omega^{p+3}} \,
\frac{\Gamma(\beta)}{\Gamma(\alpha)}\, 
\bigg\{ \mathcal{G}(\omega;0,3,3)
\nonumber \\
&
+ (\beta-\alpha) \, \bigg[ \frac{\omega}{\omega_0}\,
\mathcal{G}(\omega;0,2,2)
-\beta \, \frac{\omega}{\omega_0}\, 
\mathcal{G}(\omega;1,2,2)
- \mathcal{G}(\omega;1,3,3)\bigg]
\bigg\} \,,
\label{eq:LL_lcda}
\end{align}
where $\displaystyle p=\frac{\Gamma^{(0)}_{\rm cusp}}{2\beta_0}\ln[\alpha_{s}(\mu)/\alpha_{s}(\mu_{0})]$, the twist-3 two-particle LCDA $\phi_B^-(\omega,\mu)
=\phi_B^{-WW}(\omega,\mu)+\phi_B^{-\rm t3}(\omega,\mu)$ is a linear combination 
of the (twist-2) WW term and the genuine twist-3 term, and
\begin{align}
\mathcal{G}(\omega;l,m,n) \equiv 
G^{21}_{23}\Big(\frac{\omega}{\omega_0}\,
\Big|\,{}^{1,\beta+l}_{p+m,\alpha,p+n}\Big)\,,
\end{align}
denotes the Meijer $\mathcal{G}$-function. The evolution factors $U_\phi(\mu,\mu_0)$ and $U^{\rm t3}_\phi(\mu,\mu_0)$ are given explicitly at one-loop order~\cite{Braun:2017liq, Braun:2015pha}
\begin{eqnarray}
U_\phi(\mu,\mu_0) &=&
\exp\,\biggl\{-\frac{\Gamma^{(0)}_{\rm cusp}}{4\beta_0^2}\bigg(
\frac{4\pi}{\alpha_s(\mu_0)}\left[\ln r-1+\frac1r\right]
\nonumber\\[0.2cm]
&& \hspace*{-1.5cm}
-\frac{\beta_1}{2\beta_0}\ln^2r+\left(\frac{\Gamma^{(1)}_{\rm cusp}}{\Gamma^{(0)}_{\rm cusp}}-\frac{\beta_1}{\beta_0}\right)[r-1-\ln r]\bigg)\biggr\}
\,\left(e^{2 \gamma _E}\mu_0\right)^{\frac{\Gamma^{(0)}_{\rm cusp}}{2\beta_0}\ln r}\, r^{\frac{\gamma_{\rm t2}^{(0)}}{2\beta_0}}\,,\nonumber\\
U_\phi^{\rm t3}(\mu,\mu_0) &=& U_\phi(\mu,\mu_0)\bigg|_{\gamma_{\rm t2}^{(0)}\to \gamma_{\rm t2}^{(0)} + \gamma_{\rm t3}^{(0)}}\, ,
\end{eqnarray}
where $r = {\alpha_s(\mu)}/{\alpha_s(\mu_0)}$, $\Gamma^{(i)}_{\rm cusp}$ are the cusp anomalous dimensions at various orders and
\begin{align}
&\gamma^{(0)}_{\rm t2} = - 2 C_F\,,
&&
\gamma^{(0)}_{\rm t3} = 2N_c \, .
\end{align}
Both evolution factors satisfy the boundary condition at the reference scale $\mu_0$
\begin{align}
&U_\phi(\mu_0,\mu_0)=1\,,
&&
U^{\rm t3}_\phi(\mu_0,\mu_0)=1.
\end{align}

\end{document}